\documentclass[sigconf]{acmart}
\AtBeginDocument{%
  }

\setcopyright{acmlicensed}
\copyrightyear{2018}
\acmYear{2018}
\acmDOI{XXXXXXX.XXXXXXX}
\acmConference[Conference acronym 'XX]{Make sure to enter the correct
  conference title from your rights confirmation email}{June 03--05,
  2018}{Woodstock, NY}

\acmSubmissionID{1876}

\usepackage{amsthm}
\newtheorem{example}{Example}[section]

\usepackage[ruled,vlined]{algorithm2e}
\usepackage{graphicx}   
\usepackage{subcaption} 
\usepackage{listings}
\lstdefinestyle{promptstyle}{
    language={}, 
    basicstyle=\ttfamily\small, 
    breaklines=true, 
    columns=fullflexible, 
    frame=single, 
    frameround=tttt, 
    backgroundcolor=\color[rgb]{0.95,0.95,0.95}, 
    numbersep=5pt, 
    showstringspaces=false, 
    rulecolor=\color{black},
}

\usepackage{caption}
\DeclareCaptionType{promptblock}[Prompt Block]

\DeclareCaptionType{queryblock}[Query Block]

\usepackage{booktabs}
\usepackage{multirow}
\usepackage{balance} 

\newcommand{\newresult}[1]{\textcolor{black}{#1}}

\begin{document}

\title{Larch: Learned Query Optimization for Semantic Predicates}

\author{Fuheng Zhao}
\authornote{Equal contribution.}
\authornote{Work done while at Snowflake.}
\affiliation{%
  \institution{University of utah}
  \country{USA}
}
\email{fuheng.zhao@utah.edu}

\author{Paweł Liskowski}
\authornotemark[1]
\affiliation{%
  \institution{Snowflake Inc.}
  \country{Poland}
}
\email{pawel.liskowski@snowflake.com}

\author{Zihan Li}
\affiliation{%
  \institution{Snowflake Inc.}
  \country{USA}
}
\email{zihan.li@snowflake.com}

\author{Benjamin Han}
\authornotemark[2]
\affiliation{%
  \institution{Andreessen Horowitz}
  \country{USA}
}
\email{bhan@a16z.com}

\author{Puxuan Yu}
\affiliation{%
  \institution{Snowflake Inc.}
  \country{USA}
}
\email{puxuan.yu@snowflake.com}

\author{Varich Boonsanong}
\affiliation{%
  \institution{Snowflake Inc.}
  \country{USA}
}
\email{varich.boonsanong@snowflake.com}

\author{Dimitris Tsirogiannis}
\affiliation{%
  \institution{Snowflake Inc.}
  \country{USA}
}
\email{d.tsirogiannis@snowflake.com}

\author{Anupam Datta}
\affiliation{%
  \institution{Snowflake Inc.}
  \country{USA}
}
\email{anupam.datta@snowflake.com}


\begin{abstract}
With the advent of Large Language Models (LLMs), many database systems introduced semantic operators that enabled analytical queries over unstructured data (e.g. text, images, videos). Semantic operators typically incur high inference costs and latencies making semantic (AI) SQL queries challenging to apply on large scale datasets. At the same time, their semantic nature leads database engines to treat them as black boxes, making AISQL queries difficult to optimize. In this paper, we introduce Larch, a framework for optimizing the execution of semantic filters in AI SQL queries. Larch was inspired by two key observations: i) the high latency of semantic operators leaves significant room for computationally-heavy runtime optimization techniques, ii) unstructured data are typically accompanied by semantic information in the form of embeddings allowing for efficient semantic comparisons between \texttt{AI\_FILTER} prompts and data values. Based on these two key observations, we present two Larch variants: \textbf{Larch-A2C} and \textbf{Larch-Sel}. Larch-A2C encodes arbitrary semantic filters expression tree using an embedding-augmented Gated Graph Neural Network and formulates the filter evaluation order as a Markov decision process. In contrast, Larch-Sel leverages a supervised learning model to predict filter selectivities, subsequently applying dynamic programming to find a near-optimal evaluation order for each input row. Evaluated across diverse real-world datasets and comprehensive synthetic workloads, both Larch variants always outperform existing semantic filter optimization techniques in terms of token usage on every dataset and workload. Our results demonstrate that Larch is robust across diverse workloads across text and image modalities, \textbf{6\%--30\% lower total token costs and $3\times$--$19\times$ lower token costs overhead than Palimpzest and Quest}. 
\end{abstract}


\maketitle

\section{Introduction}
AI SQL and semantic operators have gained significant attention from both academia and industry (e.g., Snowflake, BigQuery, Microsoft) due to the promise of unlocking analytical value from vast amounts of unstructured data. An estimated 80\% of contemporary data consists of unstructured formats such as text, images, and audio~\cite{sadia2025casecomputingunstructureddata, Harbert2021}. While traditional SQL relational operators excel at structured data processing, analyzing unstructured content requires operators that understand natural language semantics. AI SQL queries interleave such semantic operators with standard relational operators~\cite{zhao2024hybridqueryingrelationaldatabases, shankar2025docetlagenticqueryrewriting, liu2025palimpzest, patel2025semanticoperatorsdeclarativemodel}. 
Semantic operators (e.g., \texttt{AI\_FILTER}, \texttt{AI\_RANK}, \texttt{AI\_AGG}) take a natural language prompt as input, which is evaluated by a Large Language Model (LLM) over one or more columns~\cite{zhao2025accesspathsefficientordering, sun2025quest, cetintemel2025makingpromptsfirstclasscitizens, trummer2025implementingsemanticjoinoperators}. These operators bridge the gap between unstructured and structured data analytics, enabling users to query unstructured content, such as filtering documents by topic, through a declarative interface.


Unlike structured data processing, where storage and CPU are the primary cost drivers, in AI SQL, the dominant bottleneck shifts to model inference costs. LLM inference credits constitute approximately 80–90\% of total query costs~\cite{liskowski2025cortexaisqlproductionsql}. Because LLMs involve billions to trillions of parameters~\cite{grattafiori2024llama3herdmodels}, each invocation also incurs seconds-level latency~\cite{fu2024serverlessllm}. 
As a result, the primary objective of optimizing these queries is to minimize LLM usage. A typical strategy is to defer the execution of semantic operators in the execution plan. The intuition is that cheaper cardinality-reducing relational operators (e.g., non-semantic predicates) should be executed first to reduce the amount of data participating in the expensive inference calls. For example, when the \texttt{WHERE} clause consists of both \texttt{AI\_FILTER}s and relational predicates, the \texttt{AI\_FILTER} operations are evaluated after the relational predicates~\cite{glenn2024blendsqlscalabledialectunifying, liskowski2025cortexaisqlproductionsql}. 
However, in real production AI SQL workloads, this simple heuristic becomes insufficient when a query contains multiple semantic filters.

Let us consider the following example of an AI SQL query that applies multiple \texttt{AI\_FILTER}s on the text column of a table:

\begin{example} 
A semantic SQL query retrieves documents from a collection that are both research papers and discuss optimization techniques in database systems.
\end{example}

\begin{center}
\begin{minipage}{0.95\linewidth}
\label{example_filter_query}
\begin{lstlisting}[style=promptstyle]
SELECT * FROM docs WHERE 
    AI_FILTER({0} is a research paper', docs.text) 
    AND AI_FILTER('{0} discusses optimization techniques in database systems', docs.text);
\end{lstlisting}
\end{minipage}
\end{center}

In this example, the query contains a conjunction of two semantic filters. A naive approach would simply evaluate the filters in the order they are written. However, to minimize expensive LLM invocations, the optimizer should execute the more selective filter first (i.e., the one more likely to return False for a conjunction, or True for a disjunction), short-circuiting evaluation of the remaining filter.
The primary difficulty in optimizing an AI SQL query containing multiple \texttt{AI\_FILTER}s in the WHERE clause is that the selectivity and correlation of these filters are highly data-dependent and unknown a priori. 
For relational predicates, optimizers rely on pre-computed metadata such as histograms or sketches to estimate selectivity~\cite{05selectivity, flajolet2007hyperloglog}. These techniques are effective because structured data exhibits well-defined value distributions.
The selectivity of an \texttt{AI\_FILTER}, however, depends entirely on latent semantic information that is only revealed at runtime. The semantic predicates consist of arbitrary natural language prompts and hence we cannot pre-compute statistics or sketches.


In Larch, we cast the problem of finding the optimal semantic filter evaluation order as an adaptive optimization problem that operates at runtime, interleaved with the actual execution of semantic operations.
Our approach builds on two key observations. First, the high latency of LLM invocations creates an unusually large window, orders of magnitude wider than in relational query processing~\cite{tang2020crocodiledb, markl2007consistent, zhu2017looking, justen2024polar}, for online runtime optimizations. Adaptive strategies, that would be impractical for relational SQL predicate, become viable now when each semantic filter evaluation takes hundreds of microseconds rather than nanoseconds. 
For instance, the system can train a lightweight model to learn correlations among filters during query execution and dynamically reorder the filters to reduce LLM inference cost. Because AND and OR are commutative, the evaluation order does not affect the query results; only the cost changes.
Second, unstructured data is frequently accompanied by pre-computed embeddings~\cite{XuSureshTang2026, pavlenco2026loadonnx}, which are orders of magnitude cheaper (100$\times$--500$\times$) than LLM invocations~\cite{OpenAI_Pricing2026}. For example, OpenAI’s in-house data agent~\cite{XuSureshTang2026} generates document embeddings during ingestion and persists them for query-time use.
Prior work treats these representations primarily as static indices for similarity search~\cite{robertson2009probabilistic, mikolov2013efficientestimationwordrepresentations, le2014distributedrepresentationssentencesdocuments}. We find that raw embedding similarity alone is too noisy to serve as a reliable selectivity signal. Instead, we re-purpose these embeddings as low-cost semantic summaries that provide the filter optimizer with a compressed representation of the input data. When used as input features for a lightweight learned model, embeddings enable accurate per-filter pass-probability predictions that drive cost-aware ordering decisions.

\newresult{Together, these observations open the door to online optimization during query execution. A natural formulation treats filter ordering as a sequential decision-making task and learns a cost-aware policy end-to-end with reinforcement learning from execution feedback, embedding features, and the expression tree. This approach is general and makes no independence assumption, but it must recover selectivity estimates, cost trade-offs, and short-circuit dynamics jointly from a single sparse reward. Our evaluation shows that it works, but that selectivity estimation dominates the remaining error (Section~\ref{sec:eval}). We therefore also explore a decomposition: a lightweight supervised model estimates per-filter selectivity from embeddings, and dynamic programming computes the evaluation order, which proves far more sample-efficient.}


To this end, we propose \textbf{Larch}, an online learning framework for \texttt{AI\_FILTER} optimization, designed to minimize inference costs without affecting output quality. Larch realizes both strategies within a shared architecture, yielding two concrete instantiations, Larch-A2C and Larch-Sel, that we evaluate against state-of-the-art methods. Our contributions are as follows:
\begin{enumerate}
    \item Exploiting the high latency of LLM inference calls, we design a pipelined online learning architecture that overlaps local model training with LLM execution. Both Larch instantiations share this design, which reclaims idle CPU cycles for model updates and hides the training overhead.

    \item We develop \textbf{Larch-A2C}, an end-to-end reinforcement learning approach that models filter ordering as a Markov decision process. A Gated Graph Neural Network encodes tree structure together with pre-computed data embeddings, and an Advantage Actor-Critic policy learns cost-aware evaluation orders directly from execution feedback. Because the policy conditions on the expression tree state, Larch-A2C can in principle capture complex semantic predicate correlations.
    

    \item Motivated by the observation that accurate per-instance selectivity estimation is the primary bottleneck, \textbf{Larch-Sel}, decomposes the problem into online selectivity estimation and exact combinatorial ordering. A lightweight neural model predicts per-filter pass probabilities from document and predicate embeddings with direct binary supervision from each LLM evaluation. A dynamic-programming solver then derives the minimum-cost evaluation sequence. 

    \item Extensive evaluations across three real-world text datasets\newresult{, one real-world image dataset,} and three semantic filter workload patterns to demonstrate that both Larch variants always outperform existing optimizers\newresult{ on every dataset and workload}, with Larch-Sel achieving the strongest results: \textbf{6\%--30\% lower total token costs and $3\times$--$19\times$ lower token costs overhead compared to state-of-the-art approaches
    }
    
\end{enumerate}

The remainder of this paper is organized as follows. Section~\ref{sec:background} provides the necessary background on semantic query optimization and focuses on optimizations for \texttt{AI\_FILTER}s. In Section~\ref{sec:larch}, we present the technical design of the Larch framework: a shared problem formulation and latency-hiding architecture (Section~\ref{sec:problem}), the end-to-end Larch-A2C (Section~\ref{sec:a2c}), and the decomposed Larch-Sel approach (Section~\ref{sec:sel}). Section~\ref{sec:eval} describes our experimental setup, where we evaluate both Larch variants against state-of-the-art approaches across multiple real-world datasets. Finally, we discuss and conclude our work in Section~\ref{sec:conclusion}.

\section{Background}
\label{sec:background}

\begin{figure*}[t] 
    \centering
    \includegraphics[width=\textwidth]{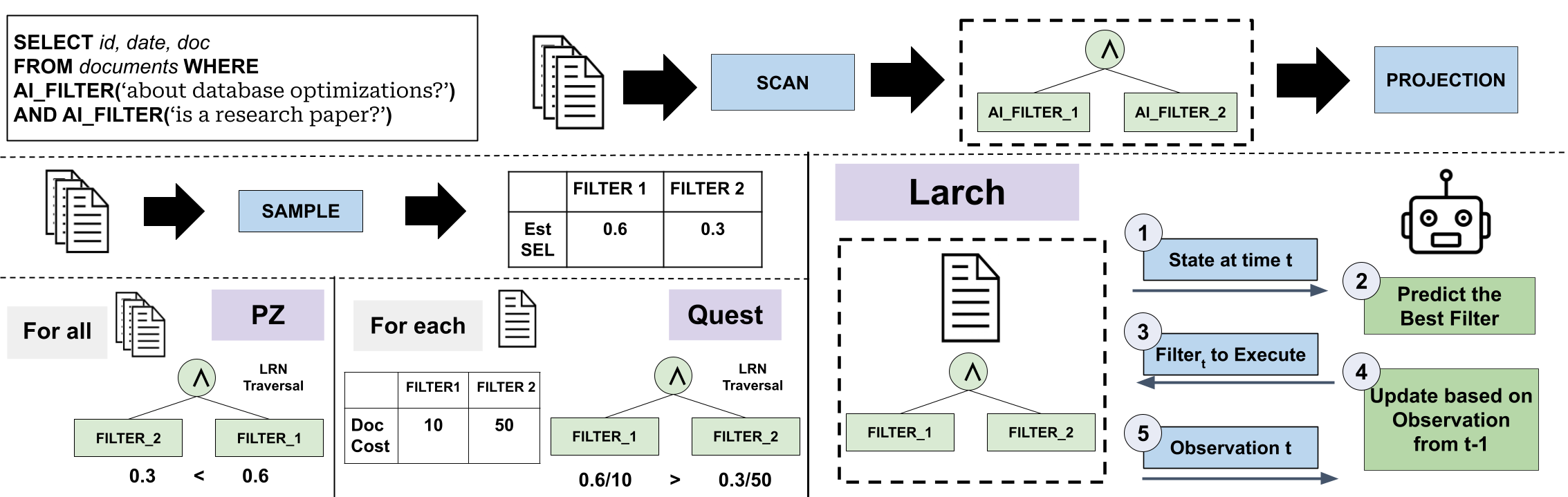}
    \caption{Overview of Larch optimizing semantic filter execution. (Top) The user issues a SQL query containing \texttt{AI\_FILTER} clauses, which is parsed into a logical execution plan involving the filter expression tree. (Bottom Left) Previous heuristic approaches (e.g., PZ and Quest) rely on global selectivity estimates based on samples to order these filters. PZ determines a static execution order for all documents, while Quest determines an order per document based on a selectivity-to-cost ratio. (Bottom Right) Larch formulates the filter ordering as an online learning process. A learned agent observes (1) the current state at time $t$, allowing it to (2) predict and (3) execute the next $filter_{t}$. At the same time, Larch (4) updates the agent using past observations to dynamically learn the optimal execution order, while (5) waiting for the next observation.}
    \label{fig:larch_arch}
\end{figure*}





AI SQL query optimization operates at two levels, logical and physical. Logical optimization refines the query plan by rewriting it with equivalent semantic operators or selecting the most suitable underlying LLM for a given task. Physical optimization targets execution efficiency, for instance by pushing relational predicates below semantic filters or by determining the evaluation order among semantic filters to minimize total inference cost.


\subsection{AI SQL Logical Optimizations}
One common strategy at the logical level is operator rewriting. Cortex AI SQL~\cite{liskowski2025cortexaisqlproductionsql} identifies opportunities to rewrite semantic primitives: when performing a semantic join (\texttt{AI\_JOIN}) on two relations $R$ and $S$, if one relation (e.g., $S$) represents a small, fixed domain of categories, the optimizer can rewrite the logical plan into an \texttt{AI\_CLASSIFY} operation. DocETL~\cite{shankar2025docetlagenticqueryrewriting} takes a different approach, introducing new operators to improve accuracy. An \texttt{AI\_FILTER} with complex logic, for instance, may be decomposed into multiple simpler \texttt{AI\_FILTER}s. When inputs are long, some systems~\cite{lin2024accurateefficientdocumentanalytics} fragment each document into sections and evaluate a chain of disjunctive \texttt{AI\_FILTER}s over the fragments.

A separate body of work uses model cascades to trade off cost and accuracy~\cite{kang2022approximateselectionguaranteesusing, patel2025semanticoperatorsdeclarativemodel, kossmann2024cascadeserveunlockingmodelcascades, chen2023frugalgptuselargelanguage, liskowski2026streamingmodelcascades}. 
\newresult{A lightweight proxy model resolves high-confidence rows and forwards only uncertain ones to the full-cost LLM, trading accuracy for cost, which may be undesirable when every row must be judged by the designated model.
Larch is orthogonal to these logical-level techniques: it does not change which model evaluates a row but decides the per-row evaluation order of multiple \texttt{AI\_FILTER}s, so it reduces cost without an accuracy trade-off.}



\subsection{Semantic Filters Execution Optimizations} 

Execution optimization focuses on the ordering of semantic operators to minimize LLM inference cost. Several systems~\cite{glenn2024blendsqlscalabledialectunifying, zhao2024hybridqueryingrelationaldatabases, Giannakouris2025SwellDB} push semantic filters above relational predicates in the plan, so that each semantic operator processes only rows that already satisfy cheaper relational predicates. Concurrent to our work, PLOP~\cite{plop} studies a complementary optimization that pulls semantic filters upward across relational operators such as joins and semi-joins, effectively delaying their evaluation to later points in the plan. PLOP shows that such pull-up transformations can reduce redundant LLM inference and develops a cost-based optimizer to identify when they are beneficial, balancing relational costs and semantic operator costs. 

Pulling semantic filters upward can cause multiple semantic predicates to become applicable at the same point in the query plan. More generally, queries may already contain multiple semantic predicates within the same \texttt{WHERE} clause. In both cases, this introduces an additional optimization problem: determining the order in which the semantic predicates should be evaluated. Recent work has explored adaptive planning for this problem~\cite{Dario25}. Cortex AI SQL~\cite{liskowski2025cortexaisqlproductionsql} collects runtime statistics such as selectivity and cost for each semantic filter. Palimpzest (PZ)~\cite{liu2025palimpzest} follows a sample-based approach, using 5\% of the total rows to estimate filter selectivity before execution begins. As illustrated in Figure~\ref{fig:larch_arch} (bottom left), these methods use the estimated selectivity to produce a static evaluation order applied uniformly across all rows. Quest~\cite{sun2025quest} also estimates filter selectivity from a sample but produces an instance-level ordering rather than a global one. For each input row $r$ and filter $i$, Quest defines a priority score $\frac{sel_{i}}{cost_{r,i}}$ and evaluates filters in descending order of this score.


Palimpzest and Quest rely on a single global selectivity estimate per filter, treating each filter's pass rate as a fixed probability throughout query execution. Leaf-node selectivities are estimated from random samples, while internal-node selectivities are derived under an independence assumption: $sel_{left} \times sel_{right}$ for conjunctions and $1 - (1-sel_{left})(1-sel_{right})$ for disjunctions. Real-world workloads, however, frequently exhibit concept drift and local correlations (e.g., data clustered by topic or ordered by timestamp)~\cite{avnur2000eddies}. Under such conditions, global estimates may fail to capture local data characteristics and produce suboptimal execution paths.

Beyond estimation accuracy, these systems also constrain how the plan is executed. PZ and Quest model the filter node as a boolean expression tree whose internal nodes are AND/OR operators and whose leaves are \texttt{AI\_FILTER} predicates (Figure~\ref{fig:larch_arch}). Both systems evaluate filters through a post-order (Left-Right-Node) traversal of this tree. Once the traversal order is fixed (across all rows in PZ, per row in Quest), execution cannot jump between branches or dynamically reorder evaluations.

\subsection{Adaptive Query Processing with Learning}
Adaptive processing and runtime plan reordering have a long history in relational query optimization~\cite{Trummer_2019, avnur2000eddies, deshpande2004overheadseddies, li2007adaptively, tzoumas2008reinforcement}. To amortize optimization and plan-switching overheads against nanosecond-fast relational operators, these systems typically batch model updates over thousands of tuples or rely on simple learning algorithms~\cite{avnur2000eddies}.

Larch shares the goal of adjusting query execution on the fly but operates under a fundamentally different cost structure. LLM invocations are orders of magnitude more expensive than relational operators in both latency and monetary cost; The penalty of suboptimal plans are correspondingly larger. A computationally heavier and more accurate model therefore becomes justified at runtime. At the same time, the substantial wait for each LLM response leaves local compute resources idle. Larch exploits this window to update its model between individual evaluation steps, completely hiding the training cost within the LLM inference latency.

\begin{figure}
    \centering
    \includegraphics[width=0.99\linewidth]{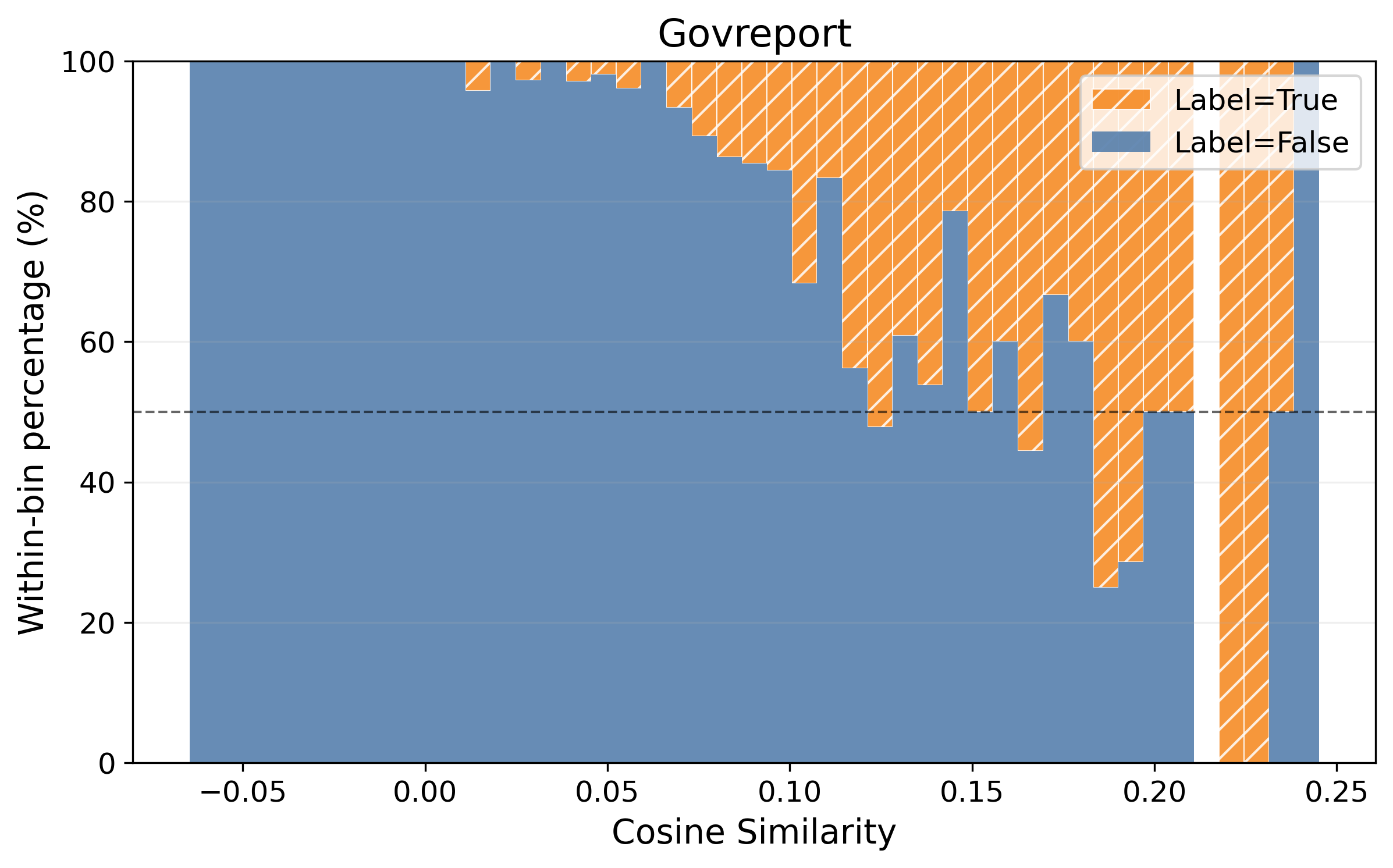}
     \caption{Fraction of True labels vs.\ cosine similarity between document and predicate prompt embeddings on GovReport (predicate: \textit{``Does the document report on military activities?''}). Higher similarity generally correlates with more True labels, but the relationship is non-monotonic and noisy.}
    \label{fig:govreport_cos_compare}
\end{figure}

\subsection{Semantic Representation}
Embedding-based representations have become widespread in AI-integrated databases~\cite{sun2025gaussdb, pan2023surveyvectordatabasemanagement}, where high-dimensional vectors capture the semantic essence of semi-structured or unstructured data~\cite{yu2024arcticembed20multilingualretrieval, voyageai_voyage3large}. For multimodal data analytics, embeddings are the new \textbf{secondary index}. 
In this work, we leverage the growing industry trend of precomputing embeddings during ingestion. OpenAI~\cite{XuSureshTang2026}, for example, generates embeddings for documents, images, and other modalities at ingestion time to support downstream querying.

A natural question arises: can the vector similarity between a pre-computed data embedding and a filter's predicate prompt serve as a low-cost proxy for estimating selectivity? We examine this relationship in Figure~\ref{fig:govreport_cos_compare} using the GovReport dataset~\cite{huang2021efficientattentionslongdocument} and a predicate from ScaleDoc~\cite{zhang2025scaledocscalingllmbasedpredicates}: \textit{``Does the document report on military activities?''}. Embeddings are generated with the \texttt{voyage} model~\cite{voyageai_voyage3large}, and True/False labels are obtained via Snowflake's \texttt{AI\_FILTER}~\cite{liskowski2025cortexaisqlproductionsql}. We compute the similarity between vectors using cosine similarity~\cite{wang2017leveraging}. As Figure~\ref{fig:govreport_cos_compare} shows, higher cosine similarity generally correlates with a higher fraction of True labels, yet the highest similarity scores correspond to a 100\% False rate, and the overall trend exhibits substantial noise.
\newresult{Embedding models capture topical overlap rather than strict predicate semantics~\cite{peng2024answerneedinstructionfollowingtext}, so raw similarity is too noisy to drive ordering on its own. Larch instead uses the embeddings as input features for an online model that learns the non-linear mapping from embeddings to LLM predicate outcomes.}


\section{Larch}
\label{sec:larch}

Larch is an online learning framework designed to optimize the execution of a filter node containing multiple \texttt{AI\_FILTER}s. A physical AI SQL query plan can have multiple such filter nodes. Given a set of semantic predicates $\mathcal{F} = \{f_1, f_2, \dots, f_n\}$ organized in a boolean expression tree, Larch determines the evaluation order of these predicates that minimizes total LLM inference cost. Consistent with prior work~\cite{liu2025palimpzest, sun2025quest}, we measure cost in terms of token usage.

As data flows through the filter node, the framework observes execution outcomes and refines its ordering decisions. We present two instantiations. \textbf{Larch-A2C} (Section~\ref{sec:a2c}) formulates filter ordering as a Markov decision process and learns an end-to-end policy via reinforcement learning, using a Gated Graph Neural Network to encode expression tree structure and data embeddings. Accurate selectivity estimation, not multi-step planning, proves to be the primary challenge in filter ordering (Section~\ref{sec:eval}). \textbf{Larch-Sel} (Section~\ref{sec:sel}) therefore takes a decomposed approach: a lightweight neural model estimates per-filter selectivities online, and exact dynamic programming derives the minimum-cost ordering over the expression tree. Both instantiations share a latency-hiding architecture (Section~\ref{sec:delayeda2c}) that overlaps local model training with remote LLM inference.


\subsection{Problem Formulation}
\label{sec:problem}

Consider a filter node whose boolean expression tree $T$ contains the $n$ predicates of $\mathcal{F}$ as leaves, connected by $\wedge$ (AND) and $\vee$ (OR) operators. For each data row $r$ entering the node, the engine must evaluate a subset of predicates sufficient to resolve $T(r)$ to a boolean value. Evaluating predicate $f_i$ on row $r$ requires an LLM inference call with token cost $c(f_i, r)$ and yields a boolean outcome. After each evaluation, the result is substituted into $T$ and the tree is reduced: an AND node with a \texttt{False} child resolves to \texttt{False}, and an OR node with a \texttt{True} child resolves to \texttt{True}, short-circuiting any remaining siblings.

The evaluation order directly determines which predicates are short-circuited and, consequently, the total cost incurred per row. Let $\sigma$ denote an adaptive ordering policy that, given the current partially evaluated tree, selects the next predicate to evaluate. The optimization objective is to find $\sigma^*$ that minimizes the expected total token cost over all rows:
$$\sigma^* = \arg\min_{\sigma} \; \mathbb{E}_{r}\!\left[\sum_{t=1}^{|\sigma(r)|} c\big(f_{\sigma_t(r)},\, r\big)\right]$$
where $|\sigma(r)|$ is the number of evaluations before the tree resolves for row $r$.

The key challenge is that predicate selectivities and correlations are unknown \emph{a priori} and may vary across the data. The objective $\sigma^*$ therefore cannot be computed ahead of time. Instead, Larch approximates it through online learning, refining its ordering decisions from execution feedback as rows are processed. The two instantiations differ in \emph{how} they use this feedback. Larch-A2C learns an end-to-end policy over the full decision sequence, while Larch-Sel estimates per-predicate statistics and delegates ordering to dynamic programming.

\subsection{Larch-A2C}
\label{sec:a2c}

Larch-A2C formulates filter ordering as a Markov Decision Process (MDP)~\cite{bellman1957markovian} and learns an evaluation policy via Advantage Actor-Critic (A2C)~\cite{sutton1998reinforcement}. The MDP is defined as follows:

\begin{itemize}
    \item \textbf{Episode ($ep$):} Each data row $r_{ep}$ entering the filter node constitutes an episode. The episode begins when $r_{ep}$ arrives and ends when the expression tree $T$ resolves to a boolean value.

    \item \textbf{State ($s_{ep,t}$):} At step $t$, the state encodes the current topology of the partially evaluated expression tree together with the semantic embeddings of remaining predicates and the input row.

    \item \textbf{Action ($a_{ep,t}$):} The agent selects one unevaluated leaf predicate from the current tree for LLM evaluation.

    \item \textbf{Transition:} The selected predicate is evaluated, its boolean result is substituted into the tree, and the tree is reduced via short-circuit logic to produce the next state $s_{ep,t+1}$.
\end{itemize}

\newresult{The action space changes at every step as predicates resolve and branches collapse. A2C handles this through two cooperating networks~\cite{a2csurvey}: the Actor produces a distribution over the current candidate predicates, and a Critic estimates the total remaining cost from the current state, so the Actor can judge each decision against a learned cost baseline.}



        

\subsubsection{State Encoding via Gated Graph Neural Network.}
The expression tree changes shape after every evaluation: resolved predicates are removed, satisfied branches are pruned, and the remaining topology shrinks. The state representation must therefore handle a variable-size, evolving graph. Larch-A2C encodes the state as a directed graph $\mathcal{G}_{ep,t} = (\mathcal{V}_{ep,t}, \mathcal{E}_{ep,t})$ that mirrors the current expression tree with bidirectional edges between each parent and its children~\cite{battaglia2018relationalinductivebiasesdeep, wu2022gnnbook}, building upon graph representation learning in query optimization~\cite{mao2019learningschedulingalgorithmsdata, DQLJOIN22}. Each leaf node representing a predicate $f_i$ carries the concatenation of document and predicate embeddings, $\mathbf{E}_{doc} \| \mathbf{E}_{filter}$. Internal nodes for $\wedge$ and $\vee$ operators carry learnable embeddings $\mathbf{e}_{\wedge}$ and $\mathbf{e}_{\vee}$. A shared projection $\mathbf{W}_{proj}$ maps all node features into a common hidden space.

To propagate structural information, we apply $K$ rounds of \emph{operator-aware} message passing. The distinction between AND and OR edges matters because short-circuit semantics differ: under AND, a single False child resolves the entire conjunction, while under OR, a single True child resolves the disjunction. Separate weight matrices for AND and OR edges allow the network to learn these distinct cost dynamics:
$$\mathbf{m}_{u \to v}^{(k)} = \mathbf{W}_{msg}^{r(u,v)} \mathbf{h}_{u}^{(k-1)}, \qquad \mathbf{h}_{v}^{(k)} = \text{GRU}\!\left( \textstyle\sum_{u \in \mathcal{N}(v)} \mathbf{m}_{u \to v}^{(k)},\; \mathbf{h}_{v}^{(k-1)} \right),$$
where $r(u,v) \in \{\wedge, \vee\}$ labels each edge with its operator type. GRU denotes a Gated Recurrent Unit~\cite{chung2014empiricalevaluationgatedrecurrent}. After $K$ rounds, mean pooling yields a global tree summary:
$$\mathbf{h}_{\mathcal{G}} = \frac{1}{|\mathcal{V}|} \sum_{v \in \mathcal{V}} \mathbf{h}_{v}^{(K)}.$$

\subsubsection{Policy and Value Networks.}
For each candidate leaf $f_i$, the Actor receives the concatenation of the leaf's own embedding $\mathbf{h}_{f_i}^{(K)}$ (enriched by $K$ rounds of message passing with its structural neighbors) and the global tree context $\mathbf{h}_{\mathcal{G}}$. A two-layer MLP scores each candidate, and a softmax restricted to the current set of unevaluated leaves produces the policy:
$$ \pi_\theta(f_i \mid s_{ep,t}) = \frac{\exp\!\big(\text{MLP}_{actor}([\mathbf{h}_{f_i}^{(K)} \| \mathbf{h}_{\mathcal{G}}])\big)}{\sum_{f_j \in \mathcal{A}_{ep,t}} \exp\!\big(\text{MLP}_{actor}([\mathbf{h}_{f_j}^{(K)} \| \mathbf{h}_{\mathcal{G}}])\big)} $$
where $\mathcal{A}_{ep,t}$ is the set of unevaluated leaves. Actions are sampled from $\pi_\theta$ for on-policy exploration. The policy's stochasticity diminishes as it converges.

The Critic estimates the expected remaining cost from the current state, $V_\phi(s_{ep,t})$, by passing the LayerNorm-normalized~\cite{ba2016layernormalization} global tree summary through a three-layer MLP. The difference between the Critic's predictions before and after each step yields the advantage, which tells the Actor whether the chosen action performed better or worse than expected:
$$\hat{A}_{ep,t} = r_{ep,t} + V_\phi(s_{ep,t+1}) - V_\phi(s_{ep,t}).$$





\subsubsection{Reward and Training.}
Each evaluation step receives reward $r_{ep,t} = -c(f_i) / C_{\text{ep,total}}$, where $c(f_i)$ denotes the token cost of evaluating predicate $f_i$ (we drop the row argument from $c(f_i, r)$ since each step processes a fixed row) and $C_{\text{ep,total}}$ sums the costs of all predicates in the initial tree. Because AI filter outputs are single-token booleans, the cost of each predicate is fully determined by the prompt (e.g., document and semantic predicate) and, similarly to PZ and Quest, $C_{\text{ep,total}}$ can be pre-computed before any evaluation begins. Normalizing by $C_{\text{ep,total}}$ makes the cumulative reward comparable across episodes with different numbers of predicates and stabilizes learning.

Since every step incurs a strictly negative reward, the episode return improves whenever a short-circuit resolves the tree early and terminates the episode before further costs accumulate. The agent therefore learns to prioritize predicates whose outcomes are most likely to trigger short-circuit resolutions.

The model is trained online via single-step temporal difference~\cite{TD1988}, minimizing:
\begin{equation*}
    \mathcal{L} = \underbrace{-\log \pi_\theta(a_t \mid s_t)\,\hat{A}_t}_{\text{policy}} \;+\; \alpha_v \underbrace{\big\| V_\phi(s_t) - y_t \big\|^2}_{\text{value}} \;-\; \beta\, \underbrace{\mathcal{H}\!\big(\pi_\theta(\cdot \mid s_t)\big)}_{\text{entropy}}
\end{equation*}
where $y_t = r_t + V_\phi(s_{t+1})$ is the TD target, $\alpha_v$ scales the value loss, and the entropy coefficient $\beta$ decays via cosine annealing to transition from exploration to exploitation~\cite{loshchilov2017sgdrstochasticgradientdescent}.

\newresult{We choose A2C with single-step TD(0) because the latency-hiding pipeline trains on one transition $(s_{t-1}, a_{t-1}, r_{t-1}, s_t)$ per LLM call. PPO~\cite{schulman2017ppo} assumes multiple epochs over thousands of transitions and overfits the 2--10-step episodes here, while multi-step returns such as GAE~\cite{schulman2016gae} chain several stale value estimates under the one-round delay and amplified error in our experiments.}

Larch-A2C provides a general end-to-end solution that makes no assumptions about predicate independence. However, the critical challenge in filter ordering is predicting per-instance selectivities accurately. If per-predicate pass probabilities are estimated accurately, the ordering step can be solved exactly rather than learned. Larch-Sel exploits this decomposition, trading the generality of the end-to-end policy for greater sample efficiency.

\subsection{Larch-Sel}
\label{sec:sel}

Larch-Sel decomposes filter ordering into two subproblems solved independently: (1)~estimating the pass probability of each predicate for the current document, and (2)~computing the minimum-cost evaluation sequence given those estimates. The decomposition rests on a key observation: if per-predicate selectivities are known, the optimal ordering over AND/OR trees can be computed exactly under an independence assumption. The learning task therefore reduces to a binary classification problem.

\subsubsection{Online Selectivity Estimation.}
For each predicate $f_i$ and document $r$, Larch-Sel maintains an online estimate of the pass probability $\hat{s}_i(r) = \Pr[f_i(r) = \texttt{True}]$ using a lightweight MLP. Document and predicate embeddings $\mathbf{E}_{doc}, \mathbf{E}_{filter} \in \mathbb{R}^{d}$ are first projected to a lower-dimensional space $\mathbb{R}^{p}$ via learned linear maps $\mathbf{W}_{doc}, \mathbf{W}_{filter}$. Let $\mathbf{d} = \mathbf{W}_{doc}\mathbf{E}_{doc}$ and $\mathbf{f} = \mathbf{W}_{filter}\mathbf{E}_{filter}$ denote the projected embeddings. The network input concatenates four components: $\mathbf{d}$ and $\mathbf{f}$ retain the individual document and predicate signals, $\mathbf{d} \odot \mathbf{f}$ captures their multiplicative interaction, and $\cos(\mathbf{d}, \mathbf{f})$ measures directional alignment independent of magnitude. In total, the feature vector has dimension $3p{+}1$:

$$\mathbf{x} = [\,\mathbf{d} \;\|\; \mathbf{f} \;\|\; \mathbf{d} \odot \mathbf{f} \;\|\; \cos(\mathbf{d},\, \mathbf{f})\,]$$

A two-layer network maps $\mathbf{x}$ to a pass probability $\hat{s}_i(r) = \sigma(\mathrm{MLP}(\mathbf{x}))$. The model is trained online with binary cross-entropy loss after each observed LLM evaluation via a single gradient step per sample. All predicates share the same network weights, so knowledge transfers across predicates from the first evaluation onward.

\subsubsection{Minimum-Cost Ordering via Dynamic Programming.}
Given the selectivity estimates $\hat{s}_i$ and token costs $c_i$, we compute the evaluation order that minimizes expected total cost under an independence assumption: each $\Pr[f_i{=}\texttt{True}]$ is treated as independent of outcomes already observed for other predicates on the same document. For flat conjunctions and disjunctions with uniform costs, the optimal ordering reduces to sorting by selectivity; Krishnamurthy et al.~\cite{krishnamurthy1986optimization} extend this to heterogeneous costs. Our DP formulation generalizes these classical results to arbitrary AND/OR trees with per-predicate costs and selectivities.

Let $T'$ denote a partially evaluated expression tree with remaining predicates $\mathcal{R}$, and let $\mathrm{OPT}(T')$ be the minimum expected cost to resolve $T'$. The optimal value satisfies the recurrence:
$$\mathrm{OPT}(T') = \min_{f_i \in \mathcal{R}} \Big[ c_i + \hat{s}_i \cdot \mathrm{OPT}(T'|_{f_i{=}\texttt{True}}) + (1 - \hat{s}_i) \cdot \mathrm{OPT}(T'|_{f_i{=}\texttt{False}}) \Big]$$
where $T'|_{f_i{=}v}$ is the tree obtained by substituting the result $v$ for $f_i$ and applying short-circuit reduction. The base case is $\mathrm{OPT}(T') = 0$ when $T'$ has resolved to a boolean. Memorization over tree structures makes the computation tractable: the number of distinct subproblems is bounded by $O(3^n)$, yielding an overall complexity of $O(n \cdot 3^n)$, where $n$ is the number of semantic predicates. For reasonably large amount of semantic predicates $n{=}10$, this amounts to roughly 590K states, and the solver runs around 20 milliseconds on a single CPU core.

At each decision point, Larch-Sel selects the predicate that achieves the minimum in the top-level recurrence. Because selectivity predictions depend on the document embedding and the model is updated after every LLM evaluation, the optimal ordering may change from one document to the next. The DP solver is therefore re-invoked per document with fresh estimates.

\subsection{Latency-Hiding Implementation}
\label{sec:delayeda2c}

Both Larch instantiations train their models online during query execution. In a production query execution engine, evaluating an \texttt{AI\_FILTER} requires an LLM inference call, which accounts for the vast majority of both total execution cost and wall-clock time~\cite{liskowski2025cortexaisqlproductionsql}. A naive synchronous implementation would stall the query engine and leave the remote LLM idle while the system waits for local model updates. To avoid this, we introduce an asynchronous pipelined architecture that overlaps local model updates with remote LLM inference.

We illustrate the pipeline using Larch-A2C, where the \textbf{delayed-update} structure is most complex. Larch-Sel follows the same three-phase pattern with its MLP gradient step replacing the actor-critic update. Consider the execution at round $t$. For brevity, we omit the episode index $ep$. The system progresses through three phases:

\begin{itemize}

    \item \textbf{Phase 1: Predict then Update.} Given the current state $s_t$, the local policy model samples the next action $a_t$ (the selected filter). Immediately after sampling, the system dispatches a background thread to train the actor-critic networks. The thread computes gradients using the completed transition $(s_{t-1}, a_{t-1}, r_{t-1}, s_t)$. The first three components are buffered from the previous round. The new state $s_t$ is strictly required here: the Critic's TD target $y_{t-1} = r_{t-1} + V_\phi(s_t)$ must bootstrap the value of the next state, which only becomes available after the tree is pruned in the \textsc{Record} phase. Phase 1 corresponds to Steps 1 and 2 in Figure~\ref{fig:larch_arch}.

    \item \textbf{Phase 2: LLM Inference.} The selected predicate is sent to the remote LLM for evaluation. A single inference call typically takes hundreds of milliseconds to seconds~\cite{fu2024serverlessllm}, orders of magnitude longer than a local gradient step. During this waiting period, the background thread launched in Phase~1 finishes updating the model weights, so the training overhead is hidden behind the LLM round-trip. Phase 2 corresponds to Steps 3 and 4 in Figure~\ref{fig:larch_arch}.

    \item \textbf{Phase 3: \textsc{Record}.} Upon receiving the boolean evaluation results from the LLM, the agent prunes the resolved branches from the expression tree to establish the next state $s_{t+1}$. The system caches the partial transition $(s_t, a_t, r_t)$ in a memory buffer. The main thread then immediately advances to round $t+1$, where $s_{t+1}$ serves as the new input state. Phase 3 corresponds to Step 5 in Figure~\ref{fig:larch_arch}.
\end{itemize}

\newresult{Model updates are thus hidden behind LLM latency, at the price of a one-round delay: the update performed during round $t+1$ uses the observations of round $t$. Updates are incremental and gradient-clipped, and the delay has minimal impact on the learned policy (Section~\ref{sec:eval}).}

\section{Experimental Evaluation}
\label{sec:eval}

\subsection{Experimental Setup}
\label{sec:experimental_setup}
We implemented Larch and the baseline methods in Python, using an evaluation framework built from two components: a query simulator and an optimization engine.
The simulator processes a collection of records (documents and images) against a specified expression tree. For each semantic predicate encountered during execution, the simulator queries a local cached \texttt{AI\_FILTER} outcomes.
The oracle outcomes in this cache are obtained by running each (document/image, semantic predicate) pair through Snowflake's \texttt{AI\_FILTER} operator~\cite{snowflake_ai_filter} .
The simulator accounts for short-circuiting in AND and OR operations, tracking token costs only for evaluated predicates. The resulting metrics thus highlight the computational savings in an optimized semantic predicate order.


In addition to Larch, the optimization engine also implements the semantic predicate evaluation logic of the algorithms listed below. 
We re-implemented the selectivity algorithms from PZ~\cite{liu2025palimpzest} and Quest~\cite{sun2025quest}. The primary comparison is between PZ, Quest, and the two proposed variants Larch-A2C and Larch-Sel.
We also include the baseline of Simple algorithm which involves no filter reordering. 
For the ablation study we additionally implement OraclePZ and OracleQuest, which bypass the sampling phase and use the dataset's ground-truth global selectivities. The oracles eliminate global-selectivity estimation error entirely and serve as a strong upper bound on what the PZ and Quest formulations can achieve.

In summary, the algorithms are as follows:
\begin{itemize}
    \item \textbf{Simple}: No filter reordering. Evaluates the \texttt{AI\_FILTER}s in a fixed order that is independent of any selectivity or cost estimate.

    \item \textbf{PZ}~\cite{liu2025palimpzest}: Reorders the tree based on estimated selectivity.

    \item \textbf{Quest}~\cite{sun2025quest}: Reorders the tree with $\text{priority}(f_i) = s_i / c_i$, where $s_i$ is the estimated selectivity and $c_i$ is the token cost, producing per-row orderings.

    \item \textbf{OraclePZ}: Runs PZ with the true selectivities computed over all rows (unavailable in practice).

    \item \textbf{OracleQuest}: Runs Quest with the true selectivities in place of the estimates.

    \item \textbf{Optimal}: Exhaustive enumeration of all valid orderings per row, selecting the minimum-cost ordering for each row. Serves as the lower bound.

    \item \textbf{Larch-A2C}: Adjusts the evaluation order online using an A2C agent that continuously refines its policy.

    \item \textbf{Larch-Sel}: Separates instance-wise selectivity estimation from planning. A supervised model predicts per-row selectivities and dynamic programming derives the order for each row.
\end{itemize}

\textbf{Implementation Details.}
\newresult{Following the original papers~\cite{liu2025palimpzest,sun2025quest}, we use a 5\% sample ratio for both PZ and Quest. At compile time, both randomly sample 5\% of the documents and estimate each \texttt{AI\_FILTER}'s global selectivity; because each sampled row requires an actual \texttt{AI\_FILTER} evaluation, this is an upfront LLM-token cost paid before execution begins. PZ then constructs a fixed evaluation plan from these estimates, while Quest reorders the filters per row using a priority score derived from the global selectivity estimates and the token cost; both combine subtree selectivities under the independence assumption described in Section~\ref{sec:background}.}

Larch-A2C uses a 3-layer operator-aware GGNN with hidden dimension 256. Leaf nodes are initialized from document and semantic-filter embeddings, and the actor and critic MLPs each have hidden dimension 128.
Larch-Sel's selectivity predictor is a two-layer MLP with approximately 144K trainable parameters, trained with a sigmoid output and binary cross-entropy loss. Document and filter embeddings are first projected to a 64-dimensional space. The projections are then concatenated with their element-wise product and their cosine similarity to form a 193-dimensional feature vector, which is passed through a hidden layer of size 64 with ReLU activation.
We use widely adopted defaults: a learning rate of $3 \times 10^{-4}$~\cite{karpathy2019recipe} and, for Larch-A2C, an entropy coefficient $\beta = 0.01$~\cite{stablebaselinesA2C}. All experiments are seeded for reproducibility.

\begin{table*}[ht]
    \centering
    \caption{Performance comparison across datasets and workload types. Workloads include Mix, Conjunction, and Disjunction. Next to each workload is the average selectivity (passing the semantic filter expressions). For each method, we report the total number of Calls and all tokens usage (Tok) for \texttt{AI\_FILTER}s, along with their percentage overhead relative to the Optimal lower-bound. Lower overhead is better. 
    We highlight the best result among all algorithms, excluding Optimal, in bold. 
    }
    \label{tab:main_results}
    \resizebox{\textwidth}{!}{
        \begin{tabular}{@{} ll cc cc cc cc cc cc @{}}
            \toprule
            \multirow{2}{*}{\textbf{Dataset}} & \multirow{2}{*}{\parbox{2cm}{\centering \textbf{Workload}\\(avg sel.)}} & 
            \multicolumn{2}{c}{\textbf{Simple}} & 
            \multicolumn{2}{c}{\textbf{PZ}} & 
            \multicolumn{2}{c}{\textbf{Quest}} & 
            \multicolumn{2}{c}{\textbf{Larch-A2C}} & 
            \multicolumn{2}{c}{\textbf{Larch-Sel}} & 
            \multicolumn{2}{c}{\textbf{Optimal}} \\
            
            \cmidrule(lr){3-4} \cmidrule(lr){5-6} \cmidrule(lr){7-8} 
            \cmidrule(lr){9-10} \cmidrule(lr){11-12} \cmidrule(l){13-14} 
            
            & & Calls & Tok & Calls & Tok & Calls & Tok 
            & Calls & Tok & Calls & Tok & Calls & Tok \\
            \midrule
            
            \multirow{7}{*}{GovReport} 
            
            & units    & K & M &  &  &  &  &  &  &  &\\

            & Mix (17\%) & (128.6 & 90.6) & (110.1 & 77.7) & (121.2 & 85.6) 
            & (107.1 & 75.4) & (\textbf{95.4} & \textbf{67.3}) & (90.8 & 64.0) \\

            &  & (+41.7\% & +41.5\%) & (+21.2\% & +21.3\%) & (+33.5\% & +33.6\%) & (+17.9\% & +17.8\%) & (+5.1\% & +5.1\%) & - & - \\

            & Conjunction (1\%) & (52.6 & 36.7) & (60.4 & 42.5) & (60.4 & 42.5) 
            & (50.9 & 35.8) & (\textbf{49.2} & \textbf{34.5}) & (46.0 & 32.2) \\

             &  & (+14.4\% & +14.0\%) & (+31.3\% & +31.8\%) & (+31.2\% & +31.8\%) & (+10.6\% & +11.2\%) & (+6.8\% & +7.1\%) & - & - \\

            & Disjunction (45\%) & (188.1 & 133.1) & (178.9 & 126.9) & (178.9 & 126.9)
            & (170.2 & 120.7) & (\textbf{161.0} & \textbf{114.1}) & (152.1 & 107.9) \\

            & & (+23.7\% & +23.4\%) & (+17.7\% & +17.6\%) & (+17.7\% & +17.6\%) & (+11.9\% & +11.9\%) & (+5.9\% & +5.8\%) & - & - \\

            \midrule
            
            \multirow{7}{*}{PubMed} 
             & units    & K & M &  &  &  &  &  &  &  &  \\

            & Mix (28\%) & (316.5 & 133.4) & (234.2 & 97.9) & (240.6 & 100.6) 
            & (223.0 & 93.5) & (\textbf{197.9} & \textbf{82.8}) & (186.0 & 77.7) \\

            & & (+70.1\% & +71.7\%) & (+25.9\% & +26.1\%) & (+29.3\% & +29.5\%) & (+19.9\% & +20.4\%) & (+6.4\% & +6.5\%) & - & - \\
            
            & Conjunction (3\%) & (167.4 & 71.5) & (151.3 & 63.9) & (151.3 & 63.9) 
            & (127.4 & 54.0) & (\textbf{121.9} & \textbf{51.5}) & (116.4 & 49.0) \\

            & & (+43.8\% & +45.9\%) & (+30.0\% & +30.4\%) & (+30.0\% & +30.4\%) & (+9.5\% & +10.2\%) & (+4.7\% & +5.0\%) & - & - \\
            
            & Disjunction (83\%) & (283.4 & 117.2) & (228.0 & 93.8) & (228.0 & 93.8) 
            & (209.2 & 85.9) & (\textbf{196.2} & \textbf{80.6}) & (173.2 & 71.2) \\

            & & (+63.7\% & +64.7\%) & (+31.7\% & +31.8\%) & (+31.7\% & +31.8\%) & (+20.8\% & +20.6\%) & (+13.3\% & +13.2\%) & - & - \\

            \midrule
            
            \multirow{7}{*}{BigPatent} 
             & units    & M & M &  &  &  &  &  &  &  &   \\

            & Mix (36\%) & (8.5 & 1188.5) & (6.1 & 852.1) & (6.2 & 861.0) 
            & (5.5 & 776.8) & (\textbf{4.9} & \textbf{688.1}) & (4.8 & 663.9) \\

            & & (+78.5\% & +79.0\%) & (+28.2\% & +28.3\%) & (+29.6\% & +29.7\%) & (+15.8\% & +17.0\%) & (+3.5\% & +3.6\%) & - & - \\
            
            & Conjunction (4\%) & (4.2 & 585.2) & (4.2 & 579.9) & (4.2 & 580.1) 
            & (3.3 & 464.0) & (\textbf{3.2} & \textbf{446.7}) & (3.2 & 439.4) \\

            & & (+31.6\% & +33.2\%) & (+31.1\% & +32.0\%) & (+31.1\% & +32.0\%) & (+5.0\% & +5.6\%) & (+1.4\% & +1.7\%) & - & - \\
            
            & Disjunction (89\%) & (7.9 & 1105.8) & (5.2 & 720.7) & (5.2 & 720.6) 
            & (4.3 & 604.6) & (\textbf{4.1} & \textbf{567.4}) & (3.9 & 546.4) \\

            & & (+99.7\% & +102.4\%) & (+31.1\% & +31.9\%) & (+31.1\% & +31.9\%) & (+9.6\% & +10.7\%) & (+3.5\% & +3.8\%) & - & - \\

            \midrule

            \multirow{7}{*}{\newresult{Ecommerce}}
             & \newresult{units}    & \newresult{M} & \newresult{M} &  &  &  &  &  &  &  &   \\

            & \newresult{Mix (23\%)} & \newresult{(5.8} & \newresult{147.5)} & \newresult{(4.5} & \newresult{96.3)} & \newresult{(4.7} & \newresult{103.5)} & \newresult{(4.2} & \newresult{81.3)} & \newresult{(\textbf{3.7}} & \newresult{\textbf{73.2})} & \newresult{(3.6} & \newresult{72.4)} \\

            & & \newresult{(+60.6\%} & \newresult{+103.8\%)} & \newresult{(+24.5\%} & \newresult{+33.0\%)} & \newresult{(+31.0\%} & \newresult{+43.0\%)} & \newresult{(+17.8\%} & \newresult{+12.4\%)} & \newresult{(+3.5\%} & \newresult{+1.1\%)} & \newresult{-} & \newresult{-} \\

            & \newresult{Conjunction (0.3\%)} & \newresult{(2.3} & \newresult{59.9)} & \newresult{(2.6} & \newresult{43.0)} & \newresult{(2.6} & \newresult{46.1)} & \newresult{(2.1} & \newresult{21.4)} & \newresult{(\textbf{2.0}} & \newresult{\textbf{20.3})} & \newresult{(2.0} & \newresult{19.9)} \\

            & & \newresult{(+12.6\%} & \newresult{+201.4\%)} & \newresult{(+27.3\%} & \newresult{+116.2\%)} & \newresult{(+28.0\%} & \newresult{+131.8\%)} & \newresult{(+4.2\%} & \newresult{+7.4\%)} & \newresult{(+0.8\%} & \newresult{+2.1\%)} & \newresult{-} & \newresult{-} \\

            & \newresult{Disjunction (67\%)} & \newresult{(8.1} & \newresult{173.0)} & \newresult{(5.9} & \newresult{128.3)} & \newresult{(6.2} & \newresult{117.5)} & \newresult{(5.4} & \newresult{102.7)} & \newresult{(\textbf{4.7}} & \newresult{\textbf{91.9})} & \newresult{(4.6} & \newresult{88.9)} \\

            & & \newresult{(+77.0\%} & \newresult{+94.6\%)} & \newresult{(+27.9\%} & \newresult{+44.3\%)} & \newresult{(+34.7\%} & \newresult{+32.2\%)} & \newresult{(+17.4\%} & \newresult{+15.5\%)} & \newresult{(+2.1\%} & \newresult{+3.4\%)} & \newresult{-} & \newresult{-} \\

            \bottomrule
        \end{tabular}
    }
\end{table*}

\textbf{Datasets.}
We evaluate on three long-document benchmark collections spanning three domains and roughly two orders of magnitude in corpus size. We also include one real-world image collection:

\begin{itemize}
    \item \textbf{GovReport}~\cite{huang2021efficientattentionslongdocument}: 973 government report summaries;
    \item \textbf{PubMed}~\cite{dernoncourt-lee-2017-pubmed}: 2{,}500 biomedical research documents;
    \item \textbf{BigPatent}~\cite{sharma2019bigpatentlargescaledatasetabstractive}: 67{,}072 patent documents\newresult{;}
    \item \textbf{Ecommerce}~\cite{lao2025sembenchbenchmarksemanticquery}: 44{,}441 product images from the SemBench e-commerce benchmark.
\end{itemize}

\textbf{Query Workloads.}
We take human-written natural-language semantic predicates from ScaleDoc~\cite{zhang2025scaledocscalingllmbasedpredicates}: each dataset is paired with a pool of 20 predicates.
From each pool we construct three workload patterns: \textit{Conjunction} (100\% $\land$), \textit{Disjunction} (100\% $\lor$), and \textit{Mixed} (50\% $\land$ / 50\% $\lor$). Relational workload studies report that 62\% of Snowflake queries contain between 3 and 10 filters~\cite{SnowWorkload2025}, so we vary the leaf count from 2 to 10 with 5 expressions per count. The resulting pool contains 45 expressions per workload pattern and hence 135 expressions per dataset. \newresult{For Ecommerce the pool also has 20 predicates: 5 are the image predicates used by SemBench's e-commerce queries Q2, Q10, and Q14~\cite{lao2025sembenchbenchmarksemanticquery} (e.g., \textit{``The image shows a (pair of) sports shoe(s) that feature the colors yellow and silver.}) and 15 are written by us in the same style (e.g., \textit{``The image shows a short-sleeved t-shirt with a graphic print.}). We build the same 135 expressions per workload pattern; the labels come from 888{,}820 \texttt{AI\_FILTER} calls, one per (image, predicate) pair.}





\textbf{Embeddings.}
The documents and queries semantic representations are produced with Voyage AI's embedding service~\cite{voyageai_voyage3large}, specifically the \texttt{voyage-multilingual-2} model invoked through Snowflake's \texttt{AI\_EMBED} function~\cite{snowflake_ai_embed}.
Document embeddings ($\mathbf{E}_{\text{doc}} \in \mathbb{R}^{1024}$) and filter-predicate embeddings ($\mathbf{E}_{\text{filter}} \in \mathbb{R}^{1024}$) are precomputed and loaded at experiment time, a widely used industry setting~\cite{XuSureshTang2026}. Both serve as input features for Larch's learning agents. \newresult{For Ecommerce, image and predicate embeddings are produced with \texttt{voyage-multimodal-3}~\cite{voyageai_voyagemultimodal3}, so Larch runs unchanged on image inputs.}

\textbf{Cost Metric.}
The primary metric is \textbf{total token cost}: the sum of tokens consumed across all LLM calls during execution. We also report the number of LLM invocations as a secondary metric. 


    

\subsection{Main Experimental Results}
\label{sec:workload_analysis}

Table~\ref{tab:main_results} reports token consumption and API calls for Larch and the baselines across the three \newresult{text datasets and the Ecommerce image dataset}, with Optimal providing the per-row lower bound. On every entry in the table, Larch-A2C uses fewer LLM calls and fewer tokens than Simple, PZ, and Quest, and Larch-Sel improves further on Larch-A2C. Larch closes most of the overhead gap to the Optimal lower bound. \newresult{Throughout this section we present two numbers: the \emph{token reduction}, i.e., the fraction of the PZ or Quest token bill that Larch saves, and the reduction in \emph{token overhead}, i.e., tokens spent above the Optimal lower bound. The first is what a user pays less; the second shows how much of the remaining headroom Larch recovers.}

Across all configurations, Larch-Sel outperforms Larch-A2C. The performance gap reflects a structural property of the problem: the filter ordering task is \emph{estimation-dominated}. Once per-instance selectivities are accurately estimated, the optimal ordering can be computed exactly under a predicate-independence assumption, leaving no residual for a learned policy to improve upon. Larch-A2C must recover selectivity estimates, cost trade-offs, and short-circuit dynamics jointly from a single reward signal over short episodes (2 to 10 steps). Larch-Sel isolates the estimation step and solves it with supervised learning (each LLM call yields a binary label), then delegates ordering to an exact DP solver, which is more sample-efficient on these workloads, where predicate correlations are moderate and the independence assumption holds well.

\newresult{\textbf{Lessons learned.}} \newresult{Filter ordering looks like a textbook reinforcement learning problem: the state is the partially evaluated expression tree, the actions are the remaining filters, the transition is the LLM's answer, and the reward is the token cost, with predicate correlations and per-row costs left for the policy to discover. That is why we built Larch-A2C first, and the formulation is viable: Larch-A2C outperforms PZ and Quest on all four datasets and all three workloads, in both LLM calls and tokens (every entry of Table~\ref{tab:main_results}). However, Larch-Sel outperforms Larch-A2C on the same twelve dataset--workload pairs. We would like to share our learned lessons from the evaluations. First, the reward carries almost no information about the action: within one row every filter costs nearly the same number of tokens (input item dominates the tokens), so the only signal that separates a good order from a bad one is how soon the episode ends, and that signal reaches the actor only through a bootstrapped critic that must itself be learned. Also, exploration in RL is not free: every suboptimal action the agent tries is a real LLM call.
Second, every LLM call returns a binary label, which is direct supervision for exactly the quantity the planner needs, and the policy-gradient formulation discards it. Third, once per-row selectivities are known the ordering problem is small enough to solve exactly, so a learned policy has to rediscover an inductive bias that the DP gets for free, and the gap between the two variants grows with the number of filters accordingly. 
The decomposition in Larch-Sel is therefore not a retreat from learning but a better placement of it: learn what is uncertain (per-row selectivity) and plan what is known (the evaluation order).}


\textbf{Performance on GovReport.}
\newresult{GovReport is the smallest collection at 973 documents, so the short horizon is a stress test for online learning. Even so, both variants use fewer LLM calls and fewer tokens than PZ and Quest on all three workloads. Larch-Sel reduces the token overhead (tokens above the Optimal lower bound) by 3--5x over PZ and Quest, and Larch-A2C by 1.4--3x. In the Conjunction workload, PZ and Quest each consume 42.5M tokens over 60.4K calls; Larch-A2C brings this to 35.8M tokens and Larch-Sel to 34.5M, cutting the token overhead from 31.8\% to 7.1\% (\textbf{a 4.5x reduction}).}

\newresult{The Conjunction case also exposes a failure mode of sampling-based baselines: at 1\% average selectivity, short-circuiting on the first filter already removes most calls, so the 5\% upfront sampling cost of PZ and Quest exceeds what reordering recovers and both end up above Simple. Larch pays no such cost because it learns from the execution itself (\S\ref{sec:update_latency}). In the Mix and Disjunction workloads, Larch-Sel lands at 5.1\% and 5.8\% token overhead, so online learning recovers a near-optimal policy with under a thousand rows.}
\newresult{In total tokens, Larch-Sel uses 10\%--19\% fewer tokens than PZ and 10\%--21\% fewer than Quest across the three workloads (12.6\% and 15.3\% summed over all 135 queries), and Larch-A2C 3\%--16\% fewer than PZ and 5\%--16\% fewer than Quest.}



\textbf{Performance on PubMed.}
\newresult{PubMed contains 2.5K biomedical documents, roughly 2.5x GovReport, and the longer horizon widens the gap to PZ and Quest. In the Mix workload, PZ and Quest incur token overheads of 26.1\% and 29.5\%; Larch-A2C brings this down to 20.4\% and Larch-Sel to 6.5\% (82.8M tokens against 97.9M for PZ). The gap is widest on the Conjunction workload: PZ and Quest both require 151.3K calls and 63.9M tokens at 30.4\% overhead, while Larch-Sel reaches 121.9K calls and 51.5M tokens at 5.0\% overhead (\textbf{a 6x reduction over PZ and Quest}). Larch-A2C sits between the two at 127.4K calls and 54.0M tokens.}
\newresult{Summed over the three workloads, Larch-Sel spends 15.9\% fewer tokens than PZ and 16.8\% fewer than Quest (14\%--19\% per workload), and Larch-A2C 8.7\% and 9.6\% fewer (4\%--16\% per workload).}


\textbf{Performance on BigPatent.}
\newresult{BigPatent is the largest corpus at 67K documents, with millions of LLM calls per workload, and Larch-Sel lands within 1.7--3.8\% of the Optimal lower bound on all three workloads. In the Mix workload, PZ and Quest consume over 850M tokens (28--29\% overhead); Larch-Sel reduces this to 688.1M tokens (3.6\%), saving 164M tokens and 1.2M LLM calls relative to PZ. The Conjunction workload exposes the limits of sampling-based planning at scale: with 4\% average selectivity, the cost of a plan depends on which filter a row fails first, which a global estimate cannot capture, so PZ and Quest settle at 32.0\% overhead ($\sim$580M tokens), barely below Simple's 33.2\%. Larch-A2C pulls the overhead down to 5.6\% and Larch-Sel to 1.7\% (446.7M tokens against the Optimal 439.4M), \textbf{a 19x reduction over PZ and Quest}. On Disjunction, PZ and Quest sit at 31.9\% overhead (720.6M tokens) while Larch-Sel reaches 3.8\% (567.4M tokens); the gap to the baselines therefore widens as the corpus grows.}
\newresult{In total tokens this is a 19\%--23\% reduction over PZ and 20\%--23\% over Quest for Larch-Sel (20.9\% and 21.3\% summed over all queries, i.e., 440--460M tokens), and 9\%--20\% for Larch-A2C.}

\begin{figure*}[t]
    \centering
    \begin{subfigure}[b]{0.24\textwidth}
        \centering
        \includegraphics[width=\textwidth]{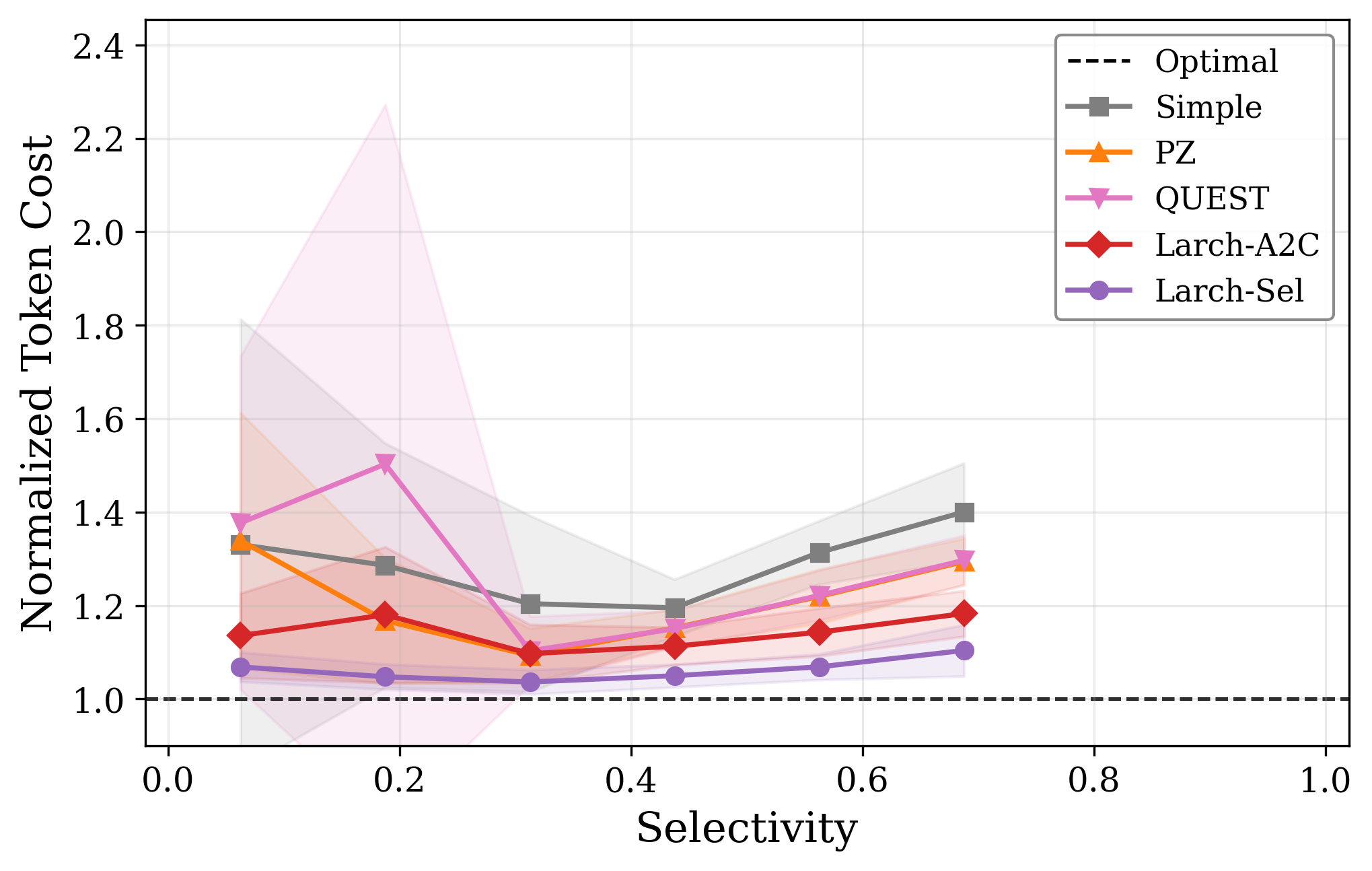}
        \caption{GovReport}
        \label{fig:SelvsTokenGovReport}
    \end{subfigure}
    \hfill
    \begin{subfigure}[b]{0.24\textwidth}
        \centering
        \includegraphics[width=\textwidth]{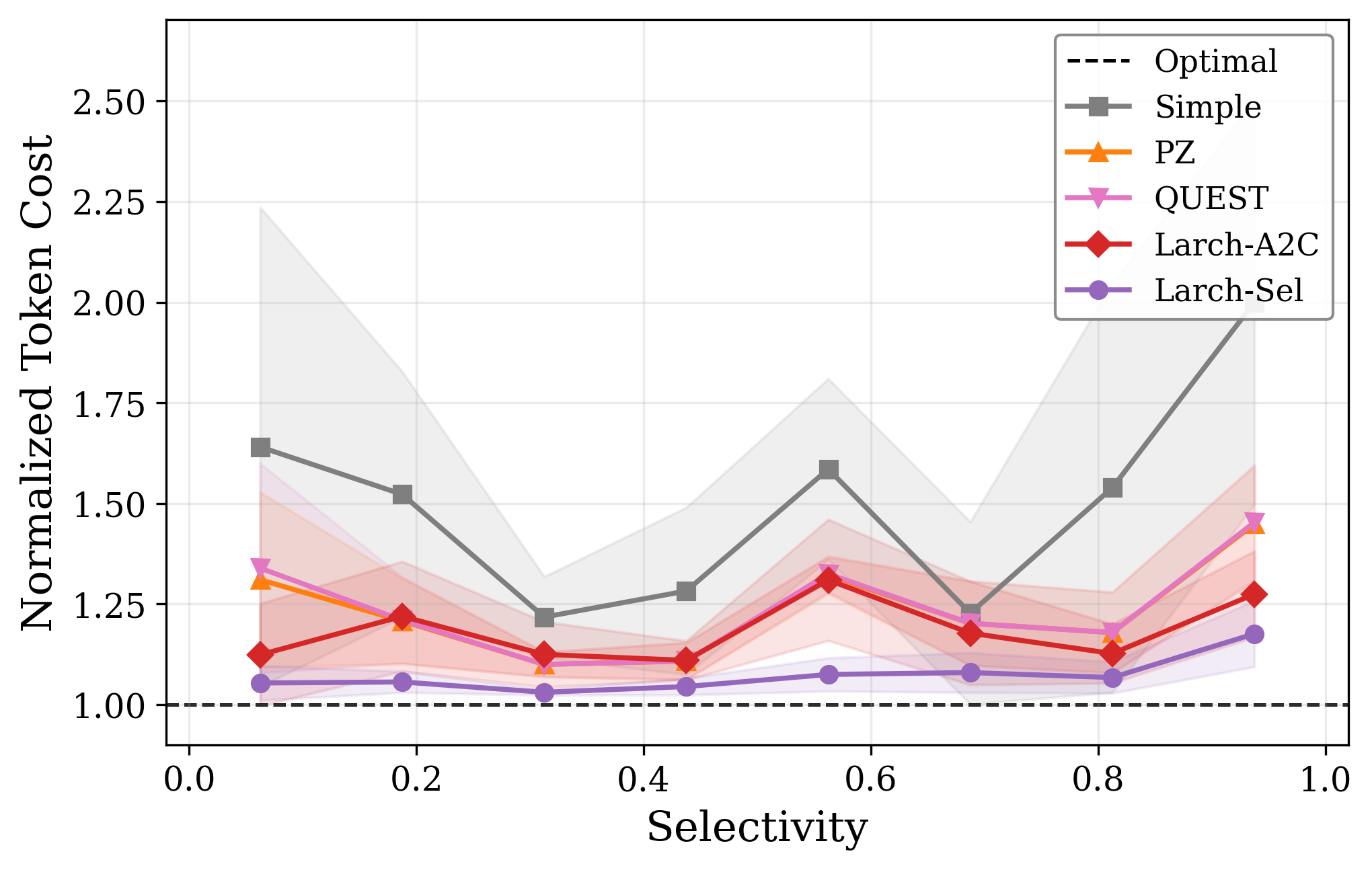}
        \caption{PubMed}
        \label{fig:SelvsTokenPubMed}
    \end{subfigure}
    \hfill
    \begin{subfigure}[b]{0.24\textwidth}
        \centering
        \includegraphics[width=\textwidth]{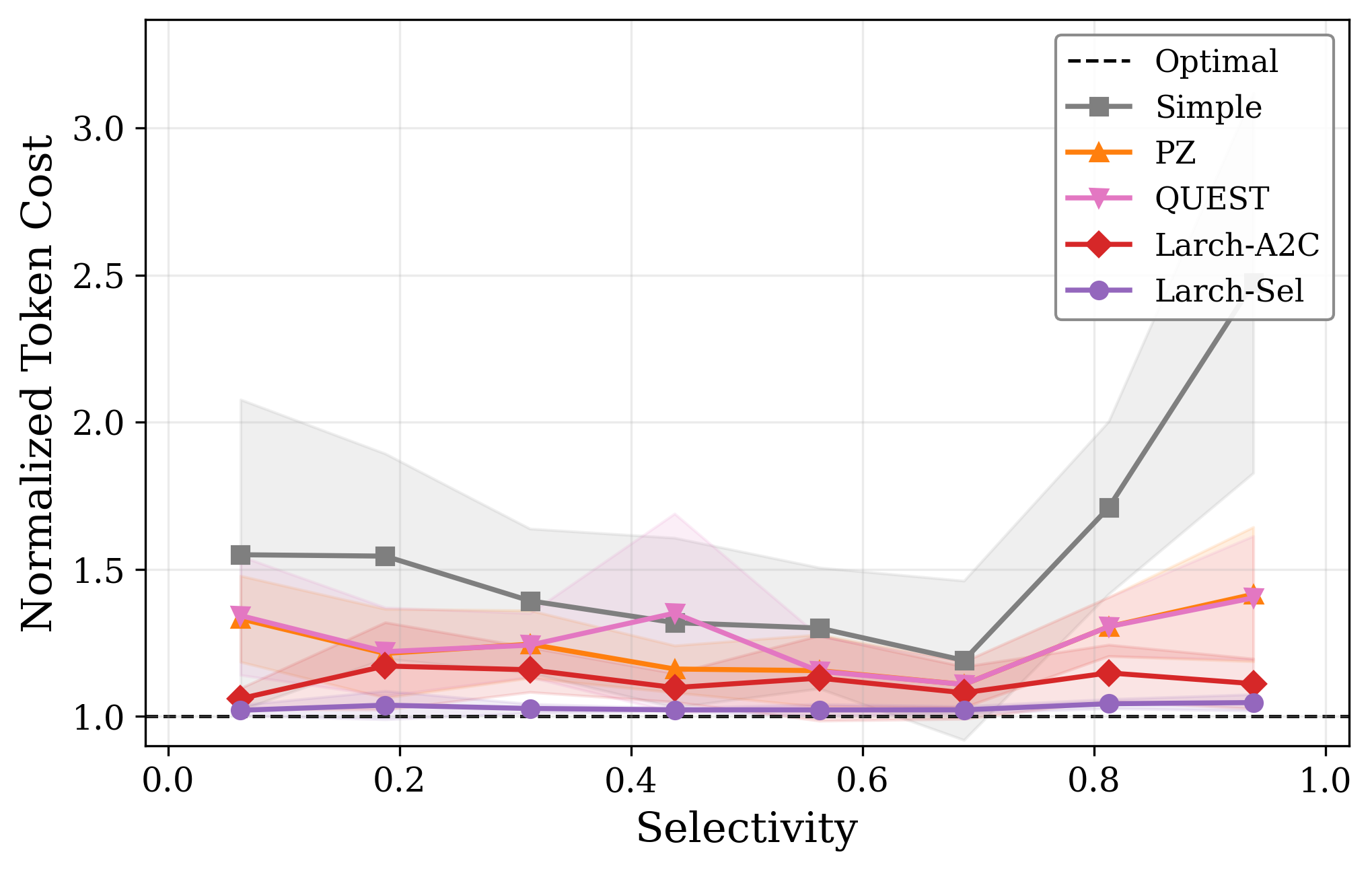}
        \caption{BigPatent}
        \label{fig:SelvsTokenBigPatent}
    \end{subfigure}
    \hfill
    \begin{subfigure}[b]{0.24\textwidth}
        \centering
        \includegraphics[width=\textwidth]{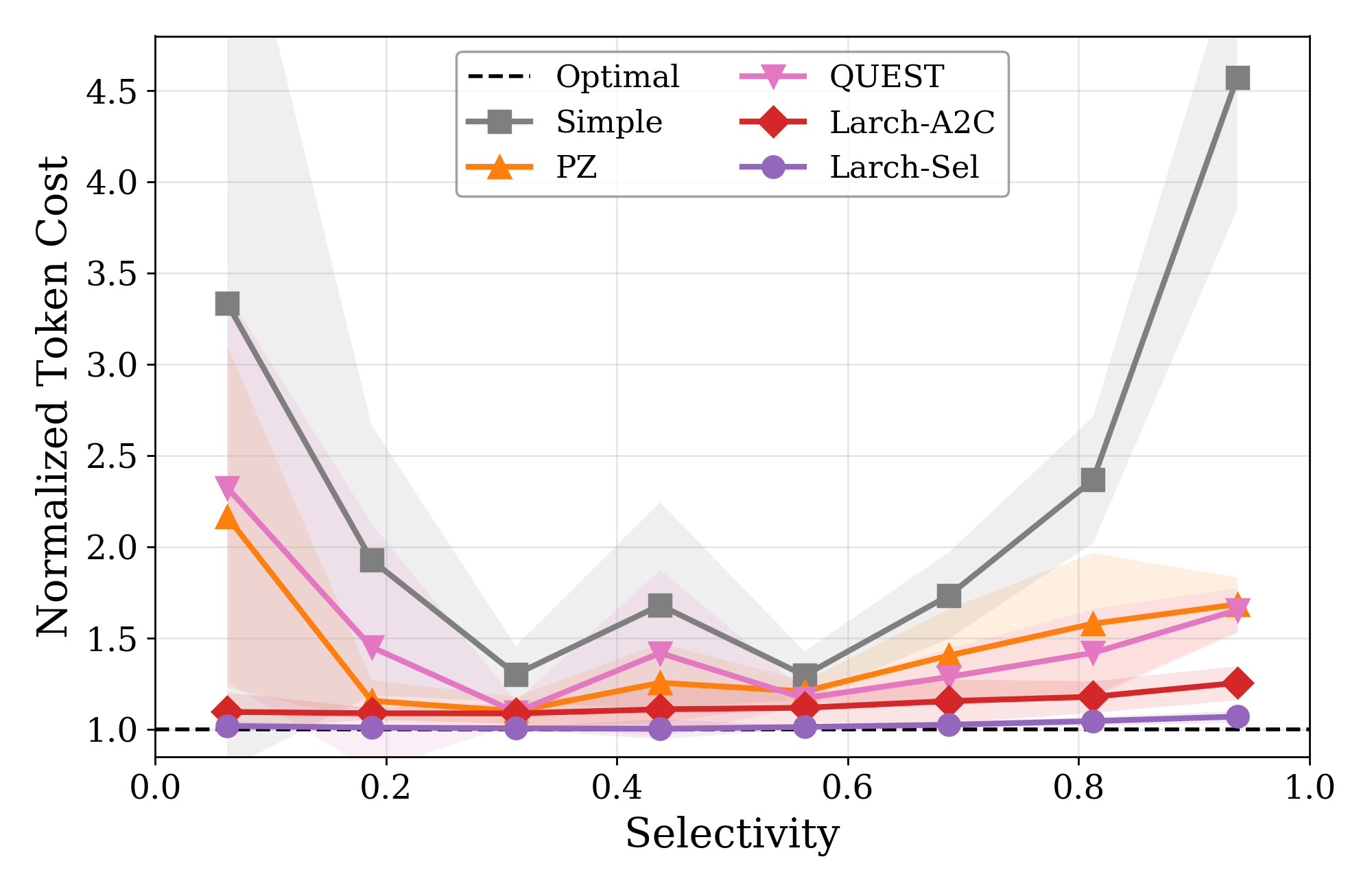}
        \caption{\newresult{Ecommerce}}
        \label{fig:SelvsTokenEcomm}
    \end{subfigure}
    
    \caption{Normalized token cost relative to the Optimal lower bound, plotted against the selectivity of the full semantic-filter expression, for (a) GovReport, (b) PubMed, \newresult{(c) BigPatent, and (d) Ecommerce}. A value of $1.0$ equals the Optimal lower bound; lower normalized token cost (y-axis) is better.}
    \label{fig:selectivity}
\end{figure*}

\newresult{\textbf{Performance on Ecommerce.}}
\newresult{Ecommerce is the image collection and, at 44K rows, the second-largest corpus; here the gap to the baselines is the widest of the four datasets. In the Mix workload, PZ and Quest consume 96.3M and 103.5M tokens (33.0\% and 43.0\% overhead). Larch-A2C reduces this to 81.3M tokens (12.4\%) and Larch-Sel to 73.2M tokens (1.1\%), saving 23--30M tokens and 0.8--1.0M LLM calls relative to the baselines, \textbf{a 29--38x reduction in token overhead}. On Conjunction (0.3\% average selectivity) the sampling-based baselines again end up above Simple in calls (2.6M versus 2.3M). Because image tokens are excluded, per-call token cost depends only on predicate length, so \emph{which} predicate is evaluated matters even at similar call counts: Simple sits at 12.6\% call overhead but 201\% token overhead, and PZ and Quest at 116\%--132\%. Larch-Sel reaches 20.3M tokens at 2.1\% overhead and Larch-A2C 21.4M at 7.4\%, a 54--61x reduction over PZ and Quest. On Disjunction, PZ and Quest incur 44.3\% and 32.2\% token overhead, Larch-A2C 15.5\%, and Larch-Sel 3.4\% (91.9M tokens against the Optimal 88.9M), saving 26--36M tokens. Across the three workloads Larch-Sel uses 24\%--53\% fewer tokens than PZ and 22\%--56\% fewer than Quest (30.7\% and 30.6\% summed over all queries), and Larch-A2C 16\%--50\% and 13\%--54\% fewer, which confirms that the embedding-driven selectivity model transfers from text to image inputs without modification.}


\begin{figure*}[htbp]
    \centering
    \begin{subfigure}[b]{0.24\textwidth}
        \centering
        \includegraphics[width=\textwidth]{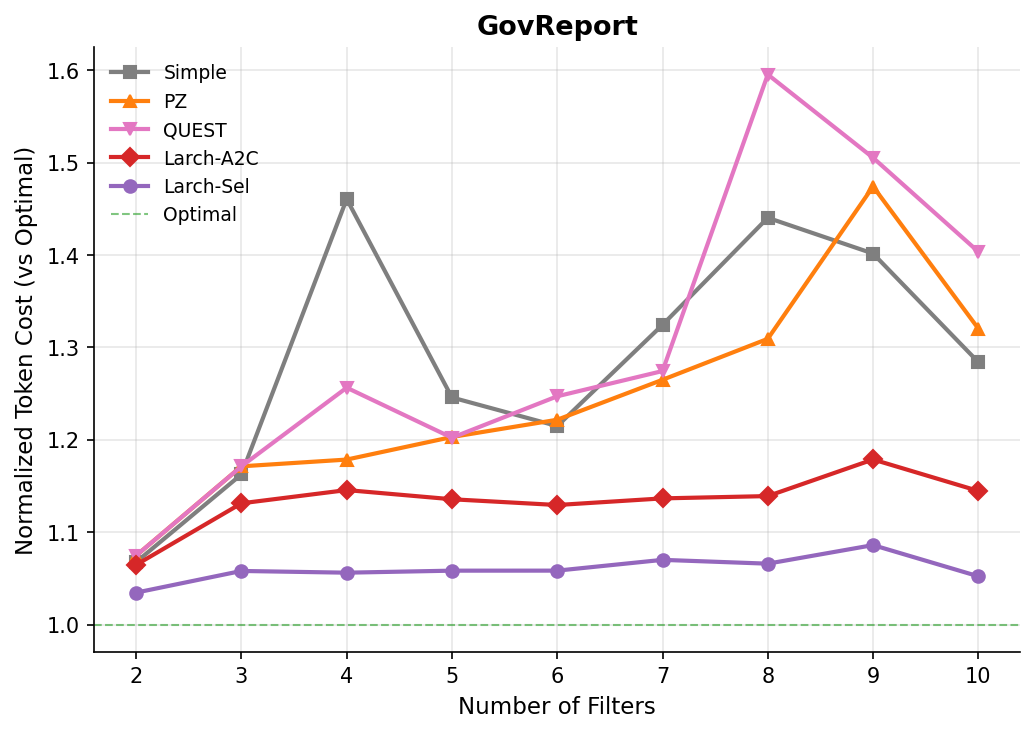}
        \caption{GovReport}
        \label{fig:numFiltervsTokenGovReport}
    \end{subfigure}
    \hfill
    \begin{subfigure}[b]{0.24\textwidth}
        \centering
        \includegraphics[width=\textwidth]{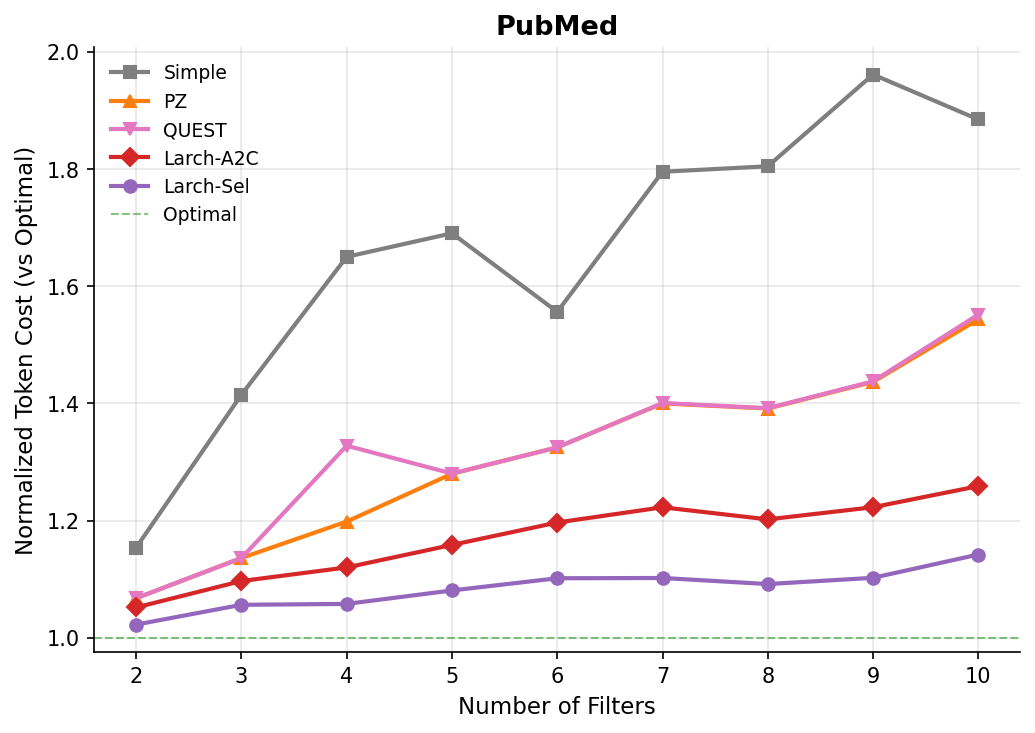}
        \caption{PubMed}
        \label{fig:numFiltervsTokenPubMed}
    \end{subfigure}
    \hfill
    \begin{subfigure}[b]{0.24\textwidth}
        \centering
        \includegraphics[width=\textwidth]{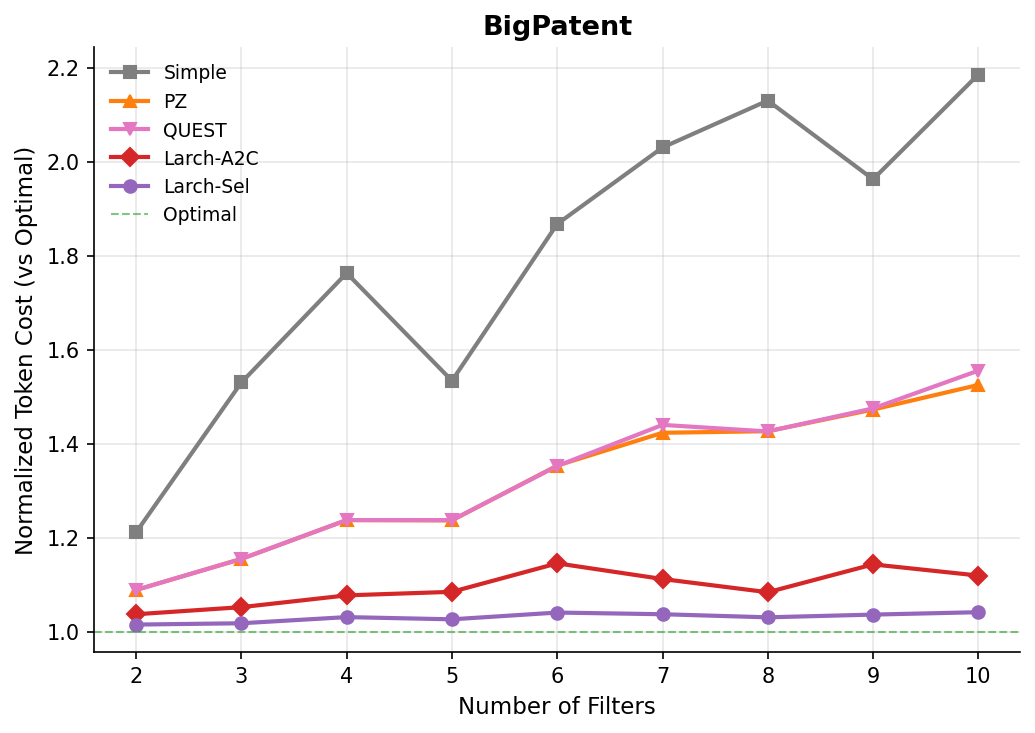}
        \caption{BigPatent}
        \label{fig:numFiltervsTokenBigPatent}
    \end{subfigure}
    \hfill
    \begin{subfigure}[b]{0.24\textwidth}
        \centering
        \includegraphics[width=\textwidth]{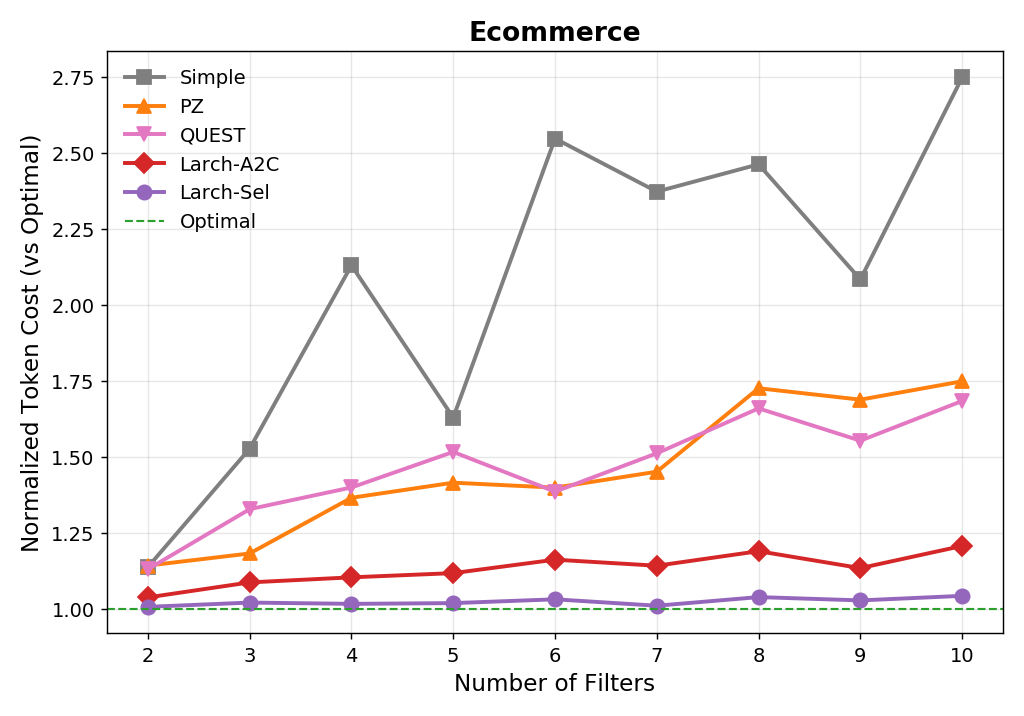}
        \caption{\newresult{Ecommerce}}
        \label{fig:numFiltervsTokenEcomm}
    \end{subfigure}
    
    \caption{Normalized token count compared to the number of semantic filters for (a) GovReport, (b) PubMed, \newresult{(c) BigPatent, and (d) Ecommerce} datasets.}
    \label{fig:vary_num_filter}
\end{figure*}

\subsection{Sensitivity to Selectivity}
\label{sec:exp_selectivity}

To characterize how each method behaves across the full selectivity range, we pool the queries from all three workloads for each dataset and group them by the selectivity of the full semantic-filter expression. Figure~\ref{fig:selectivity} plots the resulting token cost, normalized by the Optimal lower bound: a value of $1.0$ matches Optimal, and lower is better.

Larch-Sel produces the lowest normalized token cost on every selectivity bucket of all \newresult{four} datasets, and its trajectory stays flat and close to $1.0$ across the entire range. Larch-A2C is below Simple, PZ, and Quest almost everywhere, with occasional overlap with PZ or Quest near the middle of the range. Simple is the worst method by a wide margin and exceeds twice the Optimal cost in the worst buckets. 

PZ and Quest improve over Simple on most of the range but degrade at very low and very high selectivity. On GovReport (Figure~\ref{fig:SelvsTokenGovReport}), their normalized token cost spikes when the expression selectivity falls below $0.3$ or rises above $0.6$. These are the regimes in which a single global selectivity estimate is least informative: at low selectivity almost every row fails on some filter, and at high selectivity almost every row passes. The optimal ordering in each bucket depends on \emph{which} filter a given row will short-circuit on, which global estimates cannot resolve. Larch-Sel sidesteps the problem by predicting per-row selectivity, and Larch-A2C adapts its policy from online feedback.

\newresult{Ecommerce (Figure~\ref{fig:SelvsTokenEcomm}) shows the same U-shape with larger amplitude. Below selectivity $0.125$, PZ and Quest cost $2.2\times$ and $2.3\times$ Optimal while Larch-Sel stays at $1.02\times$; above $0.875$, Simple reaches $4.6\times$ and PZ and Quest $1.7\times$, while Larch-A2C and Larch-Sel remain at $1.26\times$ and $1.07\times$. The standard-deviation bands of the two Larch variants are also the narrowest, so their cost is predictable across the full selectivity range.}



\subsection{Sensitivity to the Number of Semantic Filters}
To see how each method scales with query complexity, we aggregate the results from all three workloads by the number of leaf filters in the expression (2 to 10 filters). Figure~\ref{fig:vary_num_filter} shows the resulting trajectories. Simple's curve is noisy and incurs the highest cost, exceeding $2.0\times$ Optimal at several filter counts. PZ and Quest both scale poorly: their normalized token cost grows steadily with the filter count and reaches over $1.5\times$ Optimal on PubMed and BigPatent. Their global-selectivity heuristic does not hold up as trees grow, because per-row variation in which filter short-circuits first is not captured by a single global $s_i$, and this error accumulates over more ordering decisions in the tree.

Both Larch variants scale more gracefully. Larch-A2C's overhead sits below PZ's and Quest's and is flatter on all three datasets. Larch-Sel's curve is essentially flat and stays within a narrow band around $1.0$ regardless of filter count. The reason is specific to Larch-Sel's decomposition: an accurate per-row selectivity predictor is sufficient to recover the optimal ordering under predicate independence (see \S\ref{sec:workload_analysis}), and adding more filters does not change the predictor's per-call accuracy. PZ and Quest, by contrast, rely on a single global estimate per filter, and the estimation error compounds as the number of semantic filter grows.

\newresult{The Ecommerce panel (Figure~\ref{fig:numFiltervsTokenEcomm}) is the clearest instance of this trend: PZ and Quest climb from $1.1\times$ Optimal at 2 filters to $1.7\times$ at 8--10 filters, and Simple exceeds $2.7\times$, whereas Larch-A2C stays below $1.21\times$ and Larch-Sel below $1.05\times$ at every filter count.}



\begin{table*}[h]
    \centering
    \caption{Performance comparison across datasets and workload types for Oracle PZ, Oracle Quest, A2C-Larch, and A2C-Sel. We highlight the best result among these algorithms.}
    \label{tab:oracle_vs_larch}
    \resizebox{\textwidth}{!}{
        \begin{tabular}{@{} ll cc cc cc cc @{}}
            \toprule
            \multirow{2}{*}{\textbf{Dataset}} & \multirow{2}{*}{\parbox{2cm}{\centering \textbf{Workload}\\(avg sel.)}} & 
            \multicolumn{2}{c}{\textbf{OraclePZ}} & 
            \multicolumn{2}{c}{\textbf{OracleQuest}} & 
            \multicolumn{2}{c}{\textbf{Larch-A2C}} & 
            \multicolumn{2}{c}{\textbf{Larch-Sel}} \\
            
            \cmidrule(lr){3-4} \cmidrule(lr){5-6} \cmidrule(lr){7-8} \cmidrule(l){9-10}
            
            & & Calls & Tok & Calls & Tok & Calls & Tok & Calls & Tok \\
            \midrule
\multirow{4}{*}{GovReport} 
& units    & K & M  \\ 

& Mix (17\%) & \big(99.2 (+9.3\%) & 70.0 (+9.4\%)\big) & \big(113.3 (+24.7\%) & 80.0 (+25.0\%)\big) & \big(107.1 (+17.9\%) & 75.4 (+17.8\%)\big) & \big(\textbf{95.4} (+5.1\%) & \textbf{67.3} (+5.1\%)\big) \\

 & Conjunction (1\%) & \big(\textbf{49.0} (+6.5\%) & \textbf{34.5} (+7.0\%)\big) & \big(\textbf{49.0} (+6.5\%) & \textbf{34.5} (+7.0\%)\big) & \big(50.9 (+10.6\%) & 35.9 (+11.2\%)\big) & \big(49.2 (+6.8\%) & 34.5 (+7.1\%)\big) \\
 
 & Disjunction (45\%) & \big(172.6 (+13.5\%) & 122.5 (+13.5\%)\big) & \big(172.6 (+13.5\%) & 122.5 (+13.5\%)\big) & \big(170.8 (+11.9\%) & 120.7 (+11.9\%)\big) & \big(\textbf{161.0} (+5.9\%) & \textbf{114.1} (+5.8\%)\big) \\

\midrule

\multirow{4}{*}{PubMed} 
& units    & K & M  \\ 

& Mix (28\%) & \big(210.5 (+13.1\%) & 88.0 (+13.3\%)\big) & \big(214.5 (+15.3\%) & 89.6 (+15.4\%)\big) & \big(223.0 (+19.9\%) & 93.5 (+20.4\%)\big) & \big(\textbf{197.9} (+6.4\%) & \textbf{82.8} (+6.5\%)\big) \\

 & Conjunction (3\%) & \big(123.6 (+6.2\%) & 52.4 (+6.8\%)\big) & \big(123.6 (+6.2\%) & 52.4 (+6.8\%)\big) & \big(127.4 (+9.5\%) & 54.0 (+10.2\%)\big) & \big(\textbf{121.9} (+4.7\%) & \textbf{51.5} (+5.0\%)\big) \\
 
 & Disjunction (83\%) & \big(204.2 (+17.9\%) & 83.8 (+17.7\%)\big) & \big(204.2 (+17.9\%) & 83.8 (+17.7\%)\big) & \big(209.2 (+20.8\%) & 86.1 (+20.6\%)\big) & \big(\textbf{196.2} (+13.3\%) & \textbf{80.6} (+13.2\%)\big) \\

\midrule

\multirow{4}{*}{BigPatent} 
& units    & M & M  \\ 

& Mix (36\%) & \big(5.5 (+14.7\%) & 762.1 (+14.8\%)\big) & \big(5.5 (+16.2\%) & 771.5 (+16.2\%)\big) & \big(5.5 (+15.8\%) & 776.8 (+17.0\%)\big) & \big(\textbf{4.9} (+3.5\%) & \textbf{688.1} (+3.6\%)\big) \\

 & Conjunction (4\%) & \big(3.4 (+7.9\%) & 477.0 (+8.6\%)\big) & \big(3.4 (+8.0\%) & 477.0 (+8.6\%)\big) & \big(3.3 (+5.0\%) & 464.0 (+5.6\%)\big) & \big(\textbf{3.2} (+1.4\%) & \textbf{446.7} (+1.7\%)\big) \\
 
 & Disjunction (89\%) & \big(4.5 (+13.9\%) & 625.2 (+14.4\%)\big) & \big(4.5 (+13.9\%) & 625.2 (+14.4\%)\big) & \big(4.3 (+9.6\%) & 604.6 (+10.7\%)\big) & \big(\textbf{4.1} (+3.5\%) & \textbf{567.4} (+3.8\%)\big) \\

\midrule

\multirow{4}{*}{\newresult{Ecommerce}}
& \newresult{units}    & \newresult{M} & \newresult{M}  \\

 & \newresult{Mix (23\%)} & \newresult{\big(4.0 (+12.2\%)} & \newresult{87.0 (+20.2\%)\big)} & \newresult{\big(4.3 (+20.4\%)} & \newresult{95.1 (+31.3\%)\big)} & \newresult{\big(4.2 (+17.8\%)} & \newresult{81.3 (+12.4\%)\big)} & \newresult{\big(\textbf{3.7} (+3.5\%)} & \newresult{\textbf{73.2} (+1.1\%)\big)} \\

 & \newresult{Conjunction (0.3\%)} & \newresult{\big(2.1 (+2.5\%)} & \newresult{32.6 (+63.9\%)\big)} & \newresult{\big(2.1 (+3.3\%)} & \newresult{36.4 (+83.3\%)\big)} & \newresult{\big(2.1 (+4.2\%)} & \newresult{21.4 (+7.4\%)\big)} & \newresult{\big(\textbf{2.0} (+0.8\%)} & \newresult{\textbf{20.3} (+2.1\%)\big)} \\

 & \newresult{Disjunction (67\%)} & \newresult{\big(5.6 (+20.8\%)} & \newresult{122.3 (+37.6\%)\big)} & \newresult{\big(5.9 (+28.5\%)} & \newresult{111.1 (+25.0\%)\big)} & \newresult{\big(5.4 (+17.4\%)} & \newresult{102.7 (+15.5\%)\big)} & \newresult{\big(\textbf{4.7} (+2.1\%)} & \newresult{\textbf{91.9} (+3.4\%)\big)} \\

            \bottomrule
        \end{tabular}
    }
\end{table*}

\subsection{Sensitivity to Horizon}

\begin{figure}
    \centering
    \includegraphics[width=0.8\linewidth]{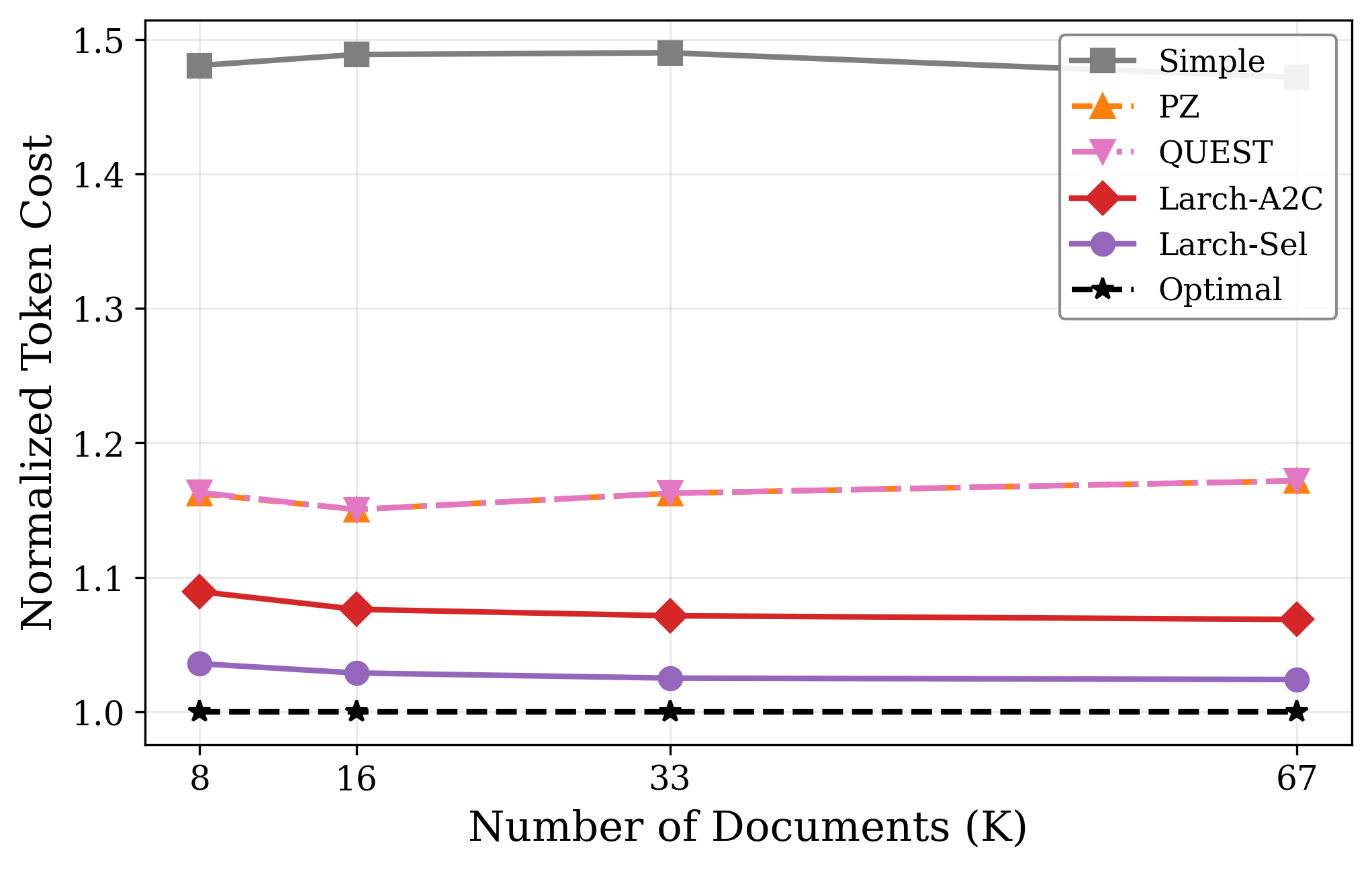}
    \caption{Normalized token cost vs. number of documents in BigPatent (summed across three workloads).}
    \label{fig:bigpatentVaryNum}
\end{figure}

To see how the horizon (the number of rows the online model observes) affects convergence, we varied the volume of BigPatent documents from 8K to 67K (Figure~\ref{fig:bigpatentVaryNum}). Throughout this range, both Larch variants consistently outperform PZ and Quest, maintaining a lower cost overhead at every interval along the curve.
The baselines exhibit relatively static performance: the Simple baseline remains flat, plateauing above $1.5\times$, while PZ and Quest show minor fluctuations around the $1.17\times$ mark. This suggests that these methods lack the adaptive mechanisms required to refine their strategies as more data becomes available.
In contrast, Larch-A2C’s normalized cost decreases steadily as the horizon extends, and Larch-Sel converges to a near-optimal $1.02\times$ cost at 67K. This trend is consistent with online learning dynamics: a longer horizon provides a richer training signal, allowing both models to converge on more efficient policies over the number of observations.

\subsection{Compare Larch to Oracle-based PZ and Quest}

To isolate the effect of estimation accuracy from the cost of estimation, we compare Larch against OraclePZ and OracleQuest. The oracles receive the true global selectivity of every filter upfront, eliminating both the global-estimation error and the 5\% sampling cost. The question is whether Larch's advantage comes from its embedding-based model or from the structural choice to route per row.

Table~\ref{tab:oracle_vs_larch} answers this: Larch-Sel outperforms both oracles on eight of the nine configurations and essentially ties on the ninth (GovReport Conjunction, $\sim$34.5M tokens: 7.0\% for the oracles, 7.1\% for Larch-Sel). On the BigPatent Mix workload, for example, OraclePZ and OracleQuest incur 14.8\% and 16.2\% overhead, while Larch-Sel reaches 3.6\%, saving 74--83M tokens over the oracles. Larch-A2C beats the oracles at BigPatent scale on the Conjunction and Disjunction workloads (5.6\% vs. 8.6\% on Conjunction, 10.7\% vs. 14.4\% on Disjunction). On the other configurations, Larch-A2C stays slightly behind the oracles while still ahead of PZ and Quest.

\newresult{Ecommerce strengthens the picture. Larch-Sel beats both oracles on all three workloads, and by a wide margin in token cost: 1.1\% overhead against 20.2\%--31.3\% on Mix, 2.1\% against 63.9\%--83.3\% on Conjunction, and 3.4\% against 25.0\%--37.6\% on Disjunction. Larch-A2C also uses fewer tokens than both oracles on every Ecommerce workload. The Conjunction row exposes why true global selectivities are not enough: OraclePZ is within 2.5\% of Optimal in calls yet 63.9\% above it in tokens, because its static plan evaluates the predicates that are likeliest to fail first regardless of their token cost, and on image inputs the per-call cost is set entirely by the predicate text. Per-row planning with learned selectivities and token costs avoids that trap.}

The result pins down the architectural gain. Even the true global selectivity $s_i$ is an average over rows, and the best static plan still picks the wrong ordering for any row whose per-row selectivity deviates from that mean. By predicting per-row selectivity directly and feeding it to the DP planner, Larch-Sel recovers the row-level variation that the global average washes out. The oracle comparison therefore reinforces the estimation-dominated view in \S\ref{sec:workload_analysis}: once per-row estimation is good enough, exact planning is straightforward.


\subsection{Larch Update and Inference Latency}
\label{sec:update_latency}

\begin{table}[t]
\centering
\caption{Average latency (ms) for Larch's inference and training steps, averaged over the three workloads for each dataset.}
\label{tab:larch-latency-breakdown}
\begin{tabular}{lcccc}
\toprule
& \multicolumn{2}{c}{\textbf{Larch-A2C}} & \multicolumn{2}{c}{\textbf{Larch-Sel}} \\
\cmidrule(r){2-3} \cmidrule(l){4-5}
\textbf{Dataset} & \textbf{Inference} & \textbf{Training} & \textbf{Inference} & \textbf{Training} \\
\midrule
GovReport & 1.59 & 9.56   & 8.51  & 6.82 \\
PubMed    & 1.69 & 10.25 & 7.96  & 6.89  \\
BigPatent & 1.76 & 10.43 & 9.75 & 6.89 \\
\bottomrule
\end{tabular}
\end{table}

The latency-hiding design works only if Larch's model update fits inside one LLM call's latency. Table~\ref{tab:larch-latency-breakdown} reports the average time Larch spends on inference (choosing the next filter to evaluate) and on training (updating the model) across the three datasets.

Larch-A2C's inference completes in under 2~ms, and its backward pass plus parameter update take 9--11~ms. The higher training cost reflects the complexity of the Gated Graph Neural Network (GGNN). Larch-Sel's training is consistent at under 7~ms, while its inference takes 8--10~ms because of the DP pass that plans the ordering from the predicted per-row selectivities.

A single \texttt{AI\_FILTER} evaluation issues an LLM call that typically takes hundreds of milliseconds~\cite{fu2024serverlessllm}. The delayed-update mechanism (\S\ref{sec:delayeda2c}) runs the training step on a background thread while the next LLM call is in flight, so the $\sim$10~ms update is fully hidden behind one LLM invocation. Inference stays on the critical path but is an order of magnitude cheaper than one LLM call, and the token savings from reordering dominate this small overhead. PZ and Quest, by contrast, pay an upfront 5\% sampling cost in LLM calls before execution begins. Larch avoids that startup cost by learning from the \texttt{AI\_FILTER} evaluations the query already performs.




\subsection{Ablation Study on Delayed Update}


\begin{table}[t]
\centering
\caption{Per-query percentage difference in total token usage with the delayed update enabled versus disabled. Values are the mean and standard deviation across 135 queries per dataset.}
\label{tab:delayed-update}
\begin{tabular}{lrr}
\toprule
\textbf{Dataset} & \textbf{Larch-A2C} & \textbf{Larch-Sel} \\
\midrule
GovReport & $-0.22\% \pm 3.67\%$ & $-0.06\% \pm 0.68\%$ \\
PubMed    & $-0.59\% \pm 4.66\%$ & $-0.06\% \pm 0.52\%$ \\
BigPatent & $+0.21\% \pm 3.96\%$ & $-0.01\% \pm 0.11\%$ \\
\bottomrule
\end{tabular}
\end{table}

The latency-hiding argument in \S\ref{sec:update_latency} assumes that deferring the gradient update by one LLM call does not noticeably degrade plan quality. Table~\ref{tab:delayed-update} tests that assumption: for each query we measure the percentage change in total token usage when the gradient update runs asynchronously on a background thread compared to a synchronous blocking update, and report the mean and standard deviation over 135 queries per dataset.

The mean change is within $\pm 0.6\%$ for every cell, so delaying the update does not cost tokens on average. The standard deviations separate the two variants: Larch-Sel stays under 0.7\% on all datasets, while Larch-A2C sits at 3.7--4.7\%. Larch-Sel is a supervised selectivity predictor whose target is row-local, so a one-step-stale gradient does not alter its predictions. Larch-A2C is an on-policy actor-critic whose gradient depends on the current rollout, so a one-step-delayed update is equivalent to training slightly off-policy. It adds per-query variance without shifting the mean, consistent with the near-zero, mixed-sign Larch-A2C means in Table~\ref{tab:delayed-update}. The delayed update is therefore safe to deploy for both variants, with the caveat that Larch-A2C shows run-to-run variation that Larch-Sel does not.
\section{Discussion and Conclusion}
\label{sec:conclusion}

In this work, we presented Larch, an online learning framework designed to optimize the execution of \texttt{AI\_FILTER}s. Larch moves away from global selectivity heuristics and instead leverages the pre-computed embeddings of unstructured data to dynamically learn instance-specific semantic predicate evaluation orders. 
We introduced two Larch variants: Larch-A2C, which formulates the problem as a Markov Decision Process, and Larch-Sel, which relies on supervised learning for instance-wise selectivity estimation. Our extensive evaluations demonstrate that while both variants outperform state-of-the-art approaches like PZ and Quest, Larch-Sel consistently achieves the best results, providing up to a $19\times$ reduction in token cost overhead compared to existing baselines.
Moreover, by learning the local semantic correlations on the fly, Larch-Sel surpasses even the Oracle approaches (OraclePZ and OracleQuest) equipped with the ground-truth selectivity as a prior.
Finally, we strategically overlap model updates with the inherently long latencies of LLM inference. With this approach, Larch delivers substantial cost savings without compromising wall-clock time on local model updates.


A key insight from our work is the difference in how these two learning paradigms handle semantic filter evaluation. The end-to-end MDP formulation (Larch-A2C) must learn selectivity estimation, cost trade-offs, and tree-aware planning jointly from a sparse reward signal. Decomposing the problem into supervised selectivity prediction and exact combinatorial ordering (Larch-Sel) proves far more sample-efficient: Larch-Sel quickly achieves near-optimal performance across our evaluation datasets (1K to 67K documents and 44K images), whereas Larch-A2C requires significantly more data to converge. \newresult{We report this as a lesson learned rather than a design choice made upfront: the sequential structure of the problem made reinforcement learning the natural first attempt, but a near-constant per-step reward and the direct supervision that every LLM call provides make the decomposed design the better use of the same execution feedback (\S\ref{sec:workload_analysis}).} Larch assumes that document embeddings are available at query time, consistent with current practice where production systems generate embeddings at ingestion~\cite{XuSureshTang2026, pavlenco2026loadonnx}. When embeddings are unavailable, the same pipeline can compute them on the fly before evaluating \texttt{AI\_FILTER} predicates. Current API pricing places embedding calls roughly 100$\times$--500$\times$ below LLM filter calls~\cite{OpenAI_Pricing2026}, so even moderate reductions in predicate evaluations usually amortize the extra embedding cost. We also note that the cost advantage narrows when \texttt{AI\_FILTER} runs on very small models with per-call costs closer to embedding models. Occasionally, upstream text-rewriting operators can also invalidate stored embeddings; re-embedding the transformed rows is then required for accurate selectivity prediction.

\newresult{Two directions remain for future work: cross-query transfer learning~\cite{zhang2023efficient, pratt1992discriminability, didrich2025learning} to mitigate the cold start, and extending Larch beyond \texttt{AI\_FILTER}s to semantic pipelines with aggregations and joins, where cross-column and cross-table semantic correlations could enable further pruning of LLM inference~\cite{madden2024databases}.}


\bibliographystyle{ACM-Reference-Format}
\bibliography{main}


\end{document}